\documentclass{emulateapj}

\shortauthors{Tran et al.}
\shorttitle{}

\begin{document}

\newcommand{\hubble}{{\it Hubble Space Telescope}}
\newcommand{\spitzer}{{\it Spitzer}}
\newcommand{\Lya}{Ly$\alpha$}
\newcommand{\Hbeta}{H$\beta$}
\newcommand{\Halpha}{H$\alpha$}
\newcommand{\Halphased}{H$\alpha_{\rm star}$}
\newcommand{\Halphagas}{H$\alpha_{\rm HII}$}
\newcommand{\NII}{\hbox{[{\rm N}\kern 0.1em{\sc ii}]}}
\newcommand{\SII}{\hbox{[{\rm S}\kern 0.1em{\sc ii}]}}
\newcommand{\OII}{\hbox{[{\rm O}\kern 0.1em{\sc ii}]}}
\newcommand{\OIII}{\hbox{[{\rm O}\kern 0.1em{\sc iii}]}}
\newcommand{\mipsmu}{$24\mu$m}
\newcommand{\iab}{$i_{\rm AB}$}
\newcommand{\bcg}{UDS40171}
\newcommand{\zcl}{$z_{\rm cl}=1.6233$}
\newcommand{\sigmacl}{$\sigma_{\rm cl}$}
\newcommand{\irc}{IRC~0218}

\newcommand{\ergu}{~$\times10^{-17}$~erg~s$^{-1}$~cm$^{-2}$}
\newcommand{\Msun}{~${\rm M}_{\odot}$}
\newcommand{\Msunyr}{~${\rm M}_{\odot}$~yr$^{-1}$}
\newcommand{\Mstar}{${\rm M}_{\star}$}
\newcommand{\logMstarMsun}{$\log({\rm M}_{\star}/{\rm M}_{\odot})$}
\newcommand{\kms}{~km~s$^{-1}$}
\newcommand{\Rproj}{R$_{\rm proj}$}

\newcommand{\zspec}{$z_{\rm spec}$}
\newcommand{\zphot}{$z_{\rm phot}$}

\newcommand{\Av}{A$_{\rm V}$}
\newcommand{\Rv}{R$_{\rm V}$}
\newcommand{\CalRv}{R$_{\rm V,SB}$}
\newcommand{\RedRv}{R$_{\rm V,Red}$}
\newcommand{\Agas}{A(\Halpha)$_{\rm HII}$}
\newcommand{\ADIMgas}{A(\Halpha)$_{\rm HII,DIM}$}
\newcommand{\Ased}{A$_{\rm V,star}$}
\newcommand{\ACalsed}{A$_{\rm V,star,SB}$}
\newcommand{\EBV}{{\rm E(B-V)}}
\newcommand{\EBVgas}{\EBV$_{\rm HII}$}
\newcommand{\EBVDIMgas}{\EBV$_{\rm HII,DIM}$}
\newcommand{\EBVsed}{\EBV$_{\rm star}$}
\newcommand{\EBVCalsed}{\EBV$_{\rm star,SB}$}

\title{ZFIRE\altaffilmark{1}: Galaxy Cluster Kinematics, \Halpha\ Star
Formation Rates, and Gas-Phase Metallicities of XMM-LSS J02182-05102 at \zcl }

\author{Kim-Vy H. Tran\altaffilmark{2}, 
Themiya Nanayakkara\altaffilmark{3},
Tiantian Yuan\altaffilmark{4},
Glenn G. Kacprzak\altaffilmark{3},
Karl Glazebrook\altaffilmark{3},
Lisa J. Kewley\altaffilmark{4},
Ivelina Momcheva\altaffilmark{5},
Casey J. Papovich\altaffilmark{2}, 
Ryan Quadri\altaffilmark{2,6},
Greg Rudnick\altaffilmark{7},
Am\'elie Saintonge\altaffilmark{8},
Lee R. Spitler\altaffilmark{9,10},
Caroline Straatman\altaffilmark{11},
Adam Tomczak\altaffilmark{2}
}

\altaffiltext{1}{http://zfire.swinburne.edu.au}
\altaffiltext{2}{George P. and Cynthia W. Mitchell Institute for
Fundamental Physics and Astronomy, Department of Physics \& Astronomy,  
Texas A\&M University, College Station, TX 77843; vy@physics.tamu.edu}
\altaffiltext{3}{Swinburne University of Technology, Hawthorn, VIC 3122, Australia}
\altaffiltext{4}{Research School of Astronomy and Astrophysics, The
Australian National University, Cotter Road, Weston Creek, ACT 2611,
Australia}
\altaffiltext{5}{Astronomy Department, Yale University, New Haven, CT 06511, USA}
\altaffiltext{6}{Mitchell Postdoctoral Fellow}
\altaffiltext{7}{Department of Physics and Astronomy, The University of Kansas, Malott Room 1082, 1251 Wescoe Hall Drive, Lawrence, KS 66045, USA}
\altaffiltext{8}{Department of Physics and Astronomy, University
College London, Gower Street, London WC1E 6BT, UK}
\altaffiltext{9}{Department of Physics and Astronomy, Faculty of Science and Engineering, Macquarie University, Sydney, NSW 2109, Australia} 
\altaffiltext{10}{Australian Astronomical Observatory, PO Box 915, North Ryde, NSW 1670, Australia}
\altaffiltext{11}{Leiden Observatory, Leiden University, P.O. Box 9513, NL-2300 RA Leiden, The Netherlands}

\begin{abstract}


We spectroscopically survey the galaxy cluster XMM-LSS J02182-05102
(hereafter \irc) using LRIS (optical) and MOSFIRE (near-infrared) on
Keck I as part of the ZFIRE survey.  \irc\ has a narrow redshift range
of $1.612<$\zspec$<1.635$ defined by 33 members of which 20 are at
\Rproj$<1$~Mpc.  The cluster redshift and velocity dispersion are
$z_{\rm cl}=1.6233\pm0.0003$ and $\sigma_{\rm cl}=254\pm50$\kms.  We
reach NIR line sensitivities of $\sim0.3$\ergu\ that, combined with
multi-wavelength photometry, provide extinction-corrected \Halpha\
star formation rates (SFR), gas phase metallicities from \NII/\Halpha,
and stellar masses.  We measure an integrated \Halpha\ SFR of
$\sim325$\Msunyr\ (26 members; \Rproj$<2$~Mpc) and show that the
elevated star formation in the cluster core (\Rproj$<0.25$~Mpc) is
driven by the concentration of star-forming members, but the {\it
average} SFR per \Halpha-detected galaxy is half that of members at
\Rproj$\sim1$~Mpc.  However, we do not detect any environmental
imprint when comparing attenuation and gas phase metallicities: the
cluster galaxies show similar trends with
\Mstar\ as to the field, e.g. more massive galaxies have larger
stellar attenuation.  \irc's gas phase metallicity-\Mstar\ relation
(MZR) is offset to lower metallicities relative to $z\sim0$ and has a
slope of $0.13\pm0.10$.  Comparing the MZR in \irc\ to the COSMOS
cluster at $z=2.1$ shows no evolution ($\Delta t\sim1$~Gyr): the MZR
for both galaxy clusters are remarkably consistent with each other and
virtually identical to several field surveys at $z\sim2$.

\end{abstract}


\section{Introduction}


The advent of sensitive infrared imaging surveys has provided a
powerful method for identifying galaxy clusters at $z>1.5$
\citep{eisenhardt:08,papovich:10,gonzalez:12,spitler:12}. We are now
able to track how clusters build up their galaxy populations over
$\sim80$\% of cosmic time ($0<z<2.1$).  Of particular interest to
current galaxy formation models are how scaling relations evolve and
depend on environment
\citep{dave:11a,dave:11b,genel:14,tonnesen:14,schaye:15,taylor:15}.
For example, the activity observed in clusters at $z>1$ combined with
the old stellar ages of cluster galaxies at $z\sim0$ indicate that
$1<z<2$ is an important epoch for understanding how star formation is
quenched \citep[e.g.][]{rudnick:12,brodwin:13}.

Essential to measuring scaling relations at $z\sim2$ is separating
first cluster galaxies from the field.  IR imaging is effective at
identifying galaxy overdensities at $z>1.5$, but broadband photometric
redshifts (\zphot) are coarse and field contamination is a concern,
especially beyond the cluster core\footnote{Although see results using
medium-band NIR filters that improve \zphot\ by a factor of $\sim5$
compared to broadband measurements \citep{whitaker:11,spitler:12}}.
This is also true for studies using narrow-band imaging to identify
\Halpha-emitting members \citep{finn:05,koyama:13a}.  Neither method
can measure cluster kinematics, and both are susceptible to
contamination by broad-line Active Galactic Nuclei (AGN).  While
\hubble\ WFC3 grism observations can measure continuum features,
e.g. the Balmer break, as well as strong emission lines
\citep{brammer:12,zeimann:13,newman:14,gobat:13}, the spectral
resolution is too low for measuring cluster kinematics and spectral
lines such as \Halpha\ and \NII\ are blended.  The WFC3 footprint also
is limited to the cluster core.

Due to these limitations, currently there are only four (proto) galaxy
clusters at $z>1.5$ that have enough members ($\gtrsim15$) for a
kinematic analysis \citep{bayliss:14,yuan:14,shimakawa:14a}.  The
large range in measured cluster velocity dispersions (550-1500\kms)
and substantial spatial substructure confirm that these are
dynamically young systems.  Because the $z\sim1.5-2$ clusters are
still assembling, we can better disentangle evolution driven by
environment versus galaxy mass
\citep{peng:10,wetzel:12,quadri:12,muzzin:12,papovich:12,bassett:13}.
For example, the discovery of intense star formation in the cluster
cores \citep{hilton:10,tran:10,zeimann:13} suggests an active
interplay between enrichment of gas in the Inter-Stellar Medium (ISM)
and the Intra-Cluster Medium (ICM).

The primary challenge to studying cluster galaxy populations at
$z>1.5$ is the need for near-IR spectroscopy.  With the installation
of efficient multi-object near-IR spectrographs such as Subaru/FMOS
\citep{kimura:10}, VLT/KMOS \citep{sharples:13}, and Keck/MOSFIRE
\citep{mclean:12}, we can now obtain key rest-frame features from
\OII$\lambda3727,3729$\AA\ to \SII$\lambda6717,6731$\AA\ for galaxy
clusters at $z>1.5$.  These spectral features have been used to
establish empirical scaling relations at $z\sim0$ that we can now test
at $z\sim2$, $i.e.$ can we detect an environmental dependence in
addition to redshift evolution?

The increasing activity in galaxy clusters with redshift raises the
question of how star formation rates (SFR), stellar masses (\Mstar),
attenuation, and gas phase metallicities are related.  Star-forming
cluster galaxies at $z>1$ define a SFR-\Mstar\ relation similar to the
field \citep[``main sequence'';][]{daddi:07,noeske:07,koyama:13b}, but
studies suggest there are environmental differences, e.g. suppressed
star formation in low-mass members and less attenuation
\citep{zeimann:13}.  If cluster galaxies follow a different relation
between specific SFR and stellar mass (SSFR-\Mstar), this could mean
that the ionization conditions are also different \citep{kewley:15}.

Recent studies of gas phase metallicity versus \Mstar\ (MZR) find that
galaxies at $z\sim2$ are uniformly offset from the local relation
\citep{erb:06a,steidel:14,sanders:15}.  However, while the MZR in the
local universe shows an environmental dependence,
\citep{cooper:08b,ellison:09,peng:14}, it is unclear if this is also
the case at $z\sim2$: several studies using stacked spectra find the
cluster galaxies to be more metal-rich than their field counterparts
\citep{kulas:13,valentino:14,shimakawa:15}, but \citet{kacprzak:15}
study $\sim50$ cluster galaxies at $z=2.1$ and find no difference from
the field.

To determine if empirical relations vary with environment at $z\sim2$,
the ZFIRE survey combines NIR spectroscopy obtained with Keck/MOSFIRE
\citep{mclean:12} with deep multi-wavelength imaging to study cluster
galaxies at $z\sim2$.  We compare to field measurements obtained by
complementary surveys such as KBSS \citep{steidel:14} and MOSDEF
\citep{kriek:15}.  Initial ZFIRE results on the COSMOS cluster at
$z=2.1$ \citep{spitler:12} include spectroscopically confirming $>50$
members and measuring the cluster's velocity dispersion
\citep{yuan:14}, measuring the cluster's gas phase metallicity-\Mstar\
relation \citep{kacprzak:15}, and investigating the ionization
properties of the Inter-Stellar Medium (ISM) of individual cluster
galaxies \citep{kewley:15}.

As part of our ZFIRE survey, we obtain optical and NIR spectroscopy of
XMM-LSS J02182-05102 (hereafter \irc)\footnote{This galaxy cluster is
also referred to as CLG0218.3-0510 by \citet{tran:10} and
\citet{santos:14}.}  at \zcl.  Originally reported by
\citet{papovich:10}, \irc\ was one of the first galaxy clusters to
show an increasing fraction of star formation with increasing local
density \citep{tran:10}.  Estimates of \irc's total star formation
rate as determined by IR observations are $>1000$\Msunyr\
\citep{popesso:12,santos:14}.  Because \irc\ is still in the process
of building its galaxy population \citep{rudnick:12}, the cluster is
an ideal target for tracking how star formation, gas phase
metallicities, and dust vary with stellar mass as a function of
environment at $z>1.5$.

In our analysis, we center \irc\ on the Brightest Cluster Galaxy (BCG)
at $(\alpha,\delta)_{\rm J2000}=(02:18:21.5,-5:10:19.9)$.  The BCG is
also currently the most distant galaxy-galaxy lens with a total mass
within the Einstein radius of $1.8\times10^{11}$\Msun\
\citep{wong:14}.  We use a Chabrier Initial Mass Function and AB
magnitudes throughout our analysis.  We assume $\Omega_{\rm m}=0.7$,
$\Omega_{\Lambda}$=0.3, and $H_0=70$\kms~Mpc$^{-1}$.  At $z=1.62$, the
angular scale is $1''=8.47$~kpc.



\section{Observations and Data}

\subsection{Target Catalog}

Spectroscopic targets were selected from the \citet{williams:09}
catalog of the Ultra-Deep Survey (UDS) taken as part of the UKIRT
Infrared Deep Sky Survey (UKIDSS), a public near-infrared imaging
survey \citep{lawrence:07}.  The $K$-selected catalogs reach
$5\sigma$-limiting magnitudes in a 1.75\arcsec~diameter of $B_{\rm
AB}<27.7$, $R_{\rm AB}<27.1$, $i_{\rm AB}<26.8$, $z_{\rm AB}<25.5$,
$J_{\rm AB}<23.9$, and $K_{\rm AB}<23.6$ mag.  We refer the reader to
\citet{williams:09} and \citet{quadri:12} for more details on the
photometric and \zphot\ catalogs used in this work.

\subsection{Optical Spectroscopy: Keck/LRIS}

Using the {\it Low Resolution Imaging Spectrometer}
\citep[LRIS;][$5.5\arcmin \times 8\arcmin$ field of view;
$0.135\arcsec$ per pixel]{oke:95} on Keck, we carried out a
spectroscopic survey centered on the cluster on 19 \& 20 October 2012
(NASA/Keck Program ID 48/2012B).  The primary LRIS targets were
selected to be brighter than \iab$=21$ mag and the secondary targets
(mask fillers) between $21<$\iab$<24$ mag.  No morphological selection
was applied.  Of the primary targets, higher priority was also given
to objects identified to be candidate star-forming cluster galaxies
\citep{tran:10}, candidate Lyman-Break Galaxies at \zphot$>1.35$, and
\OII-emitters identified from narrow-band imaging \citep{tadaki:12}.  

We use the 600/4000 grism for the blue side of LRIS
($0.38<\lambda<0.58$ $\mu$m) and the 600/10000 grating for the red
side ($0.70<\lambda<1.0$ $\mu$m); with 1\arcsec~slit widths, the
corresponding resolution is 4.0\AA\ and 4.7\AA\ respectively.
Observing conditions were excellent with median seeing of about
0.6\arcsec.  A subset of high priority targets, e.g. the brightest
cluster galaxies, were targeted in three of the four observed masks.
For masks 1, 2, and 3 which included common high priority targets, we
obtained $9\times20$ minute exposures.  For mask 4 which included a
more general redshift selection, we obtained $5\times20$ minute
exposures.  The four LRIS masks have a total of 136 targets including
repeats.  For all the exposures, we dithered by $\pm1\arcsec$ along
the slit to minimize sky residuals.  Standard calibrations including
flat-fields and arc lamps (ArCdHgNeZn) for each mask also were taken.

To reduce the spectra, we follow \citet{tran:07} and use
IRAF\footnote{IRAF is distributed by the National Optical Astronomy
Observatories, which are operated by the Association of Universities
for Research in Astronomy, Inc., under cooperative agreement with the
National Science Foundation.} routines with custom software provided
by D. Kelson \citep{kelson:98,kelson:03}.  To summarize, we first
convert the multi-extension files into single images for the blue and
red sides; we reduce the blue and red sides of LRIS separately.
Cosmic rays are particularly problematic on the red side and we remove
as many as possible using {\it crutil} in IRAF; the blue side is
minimally affected by cosmic rays.  We flat-field the science frames,
determine the vacuum wavelength solution with the arclamp, remove the
sky lines, and rectify the 2D spectra.  We median the rectified
science frames for each mask to obtain a single combined image.

The 2D spectra first were visually inspected to identify all potential
sources including emission-line only objects.  The 1D spectra were
extracted by summing the five rows centered on the source
(0.135\arcsec~per pixel) of each 2D spectrum; the wavelength coverage
for the extracted spectra is $3800-5800$\AA\ (blue side) and
$7000-10000$\AA\ (red side).  For galaxies at $z\sim1.6$, this
corresponds to rest-frame UV where the continuum is detected for only
a handful of systems.  To measure redshifts, we use {\it xcsao} in
IRAF \citep{kurtz:92} with various templates for star-forming,
quiescent, post-starburst, Lyman-break, and \Lya\ galaxies; we refer
the reader to \citet{tran:05b} for details.

\subsection{Near-Infrared Spectroscopy: Keck/MOSFIRE}

At \zcl, most of the rest-frame optical spectral features such as
\Hbeta\ and \Halpha\ have shifted to the near-IR.  To measure these
features and provide continuous spectral coverage of the cluster
galaxies, we used Keck/MOSFIRE \citep{mclean:12} on 23 \& 24 December
2013 and on 10-13 February 2014.  Primary targets were galaxies
spectroscopically confirmed with the LRIS observations to be at
$z>1.5$ (see Fig.~\ref{fig:zhist}).  We also included objects from the
3D-HST survey with grism redshifts $z>1.5$ (Momcheva et al., in prep).
Secondary targets were selected to have \zphot$>1.4$ using the
Williams catalog, and the lowest priority fillers had \zphot$<1.4$ and
\iab$<22$ mag.

We obtained YJH spectroscopy corresponding to wavelength ranges of
$0.97-1.12\mu$m, $1.15-1.35\mu$m, and $1.46-1.81\mu$m respectively.
Using $0.7''$ slit-widths (pixel scale is $0.18''$), the spectral
resolution of $R\sim3600$ corresponds to a dispersion of 1.086, 1.303,
and 1.627 \AA/pixel for YJH.  The MOSFIRE spectroscopic field of view
is $3'\times6'$, and we observed four masks with approximately 25-35
objects targeted in each to map the galaxy cluster
(Fig.~\ref{fig:zhist}).  All the masks were designed using the MAGMA
slitmask design software
\footnote{www2.keck.hawaii.edu/inst/mosfire/magma.html}. A slit star was included in each mask to monitor the flux throughput
and seeing.  Flat-fields and arcs (Neon, Argon) were taken during
afternoon calibrations for each mask.

For the December 2013 run, we designed three masks that were observed
with both J and H.  The seeing FWHM on the first night was
$\sim0.7-0.9''$ and improved to $\sim0.6''$ on the second night.
Adjusting for the seeing conditions to reach approximately the same
flux limit in all masks, we integrated for a total of $2880-3360$
seconds and $2880-5880$ seconds in the J and H bands respectively;
individual exposure times for both JH bands were 120 sec.  All
spectroscopy was taken with an ABBA dither pattern and an offset of
$2.5''$.  Long-slit spectroscopy of the standard star Feige 15 was
taken on both nights to correct for telluric absorption by the
atmosphere and to flux calibrate the data.

For the February 2014 run, we used a single mask to obtain Y band
spectroscopy of mostly confirmed cluster members.  Conditions on the
first night were poor with seeing FWHM of $1.2''$ (none of these
observations are usable) but improved to $\sim0.7-0.9''$ for the
following three nights.  The Y mask was observed each night for
$5040-6660$ sec with individual exposure times of 180 sec.  The total
integration time is 24300 sec, but we note that useful time on target
is less due to varying conditions on the first two nights.  Data taken
under poorer weather conditions were given less weight when combining
the observations, i.e. the data were weighted by the S/N of the
continuum.

To reduce the MOSFIRE spectroscopy, we use the publicly available data
reduction pipeline (DRP) developed by the instrument
team\footnote{https://github.com/Mosfire-DataReductionPipeline/MosfireDRP}.
The DRP provides background-subtracted, rectified, and (vacuum)
wavelength calibrated 2D spectra for each slit.  Using custom IDL
routines, we then correct the 2D spectroscopy for telluric absorption
and flux calibrate using the standard star observations.  To extract
the 1D spectra and associated $1\sigma$ error spectra, we fit a
Gaussian profile along the spatial direction to determine the
extraction aperture for each object; typical extraction apertures are
$\sim1''$.  For objects that are too faint to fit a spatial profile,
we use the Gaussian profile fit to the slit star in the same mask.

The 2D spectra were visually inspected to identify all potential
sources; for the most part, these detections are emission lines,
e.g. \Halpha\ in the H-band for galaxies at $z\sim1.6$.  Spectroscopic
redshifts were determined with two-step method: 1) using $xcsao$ in
IRAF with an emission-line template and, where possible, combining JH
observations, to measure a redshift and 2) using this redshift to fit
1D Gaussians to emission lines along the spectral direction and
propagating errors using the corresponding error spectrum.

For more details on our data reduction, we refer to our MOSFIRE
analysis of the COSMOS cluster \citep{yuan:14}.  Especially thorough
reviews of MOSFIRE observing strategies and data reduction are also
presented by \citet{kriek:15} and \citet{steidel:14}; note that the
latter uses the public data reduction pipeline (DRP) released for
MOSFIRE. Both publications provide extensive discussion on the
capabilities of the instrument and strategies for post-processing.
A full description of our data reduction for \irc\ and the COSMOS
galaxy cluster at $z=2.1$ also will be presented in Nanayakkara et
al. (in prep).



\subsection{Redshift Catalog}


\begin{figure*}
\centerline{\includegraphics[width=0.8\textwidth]{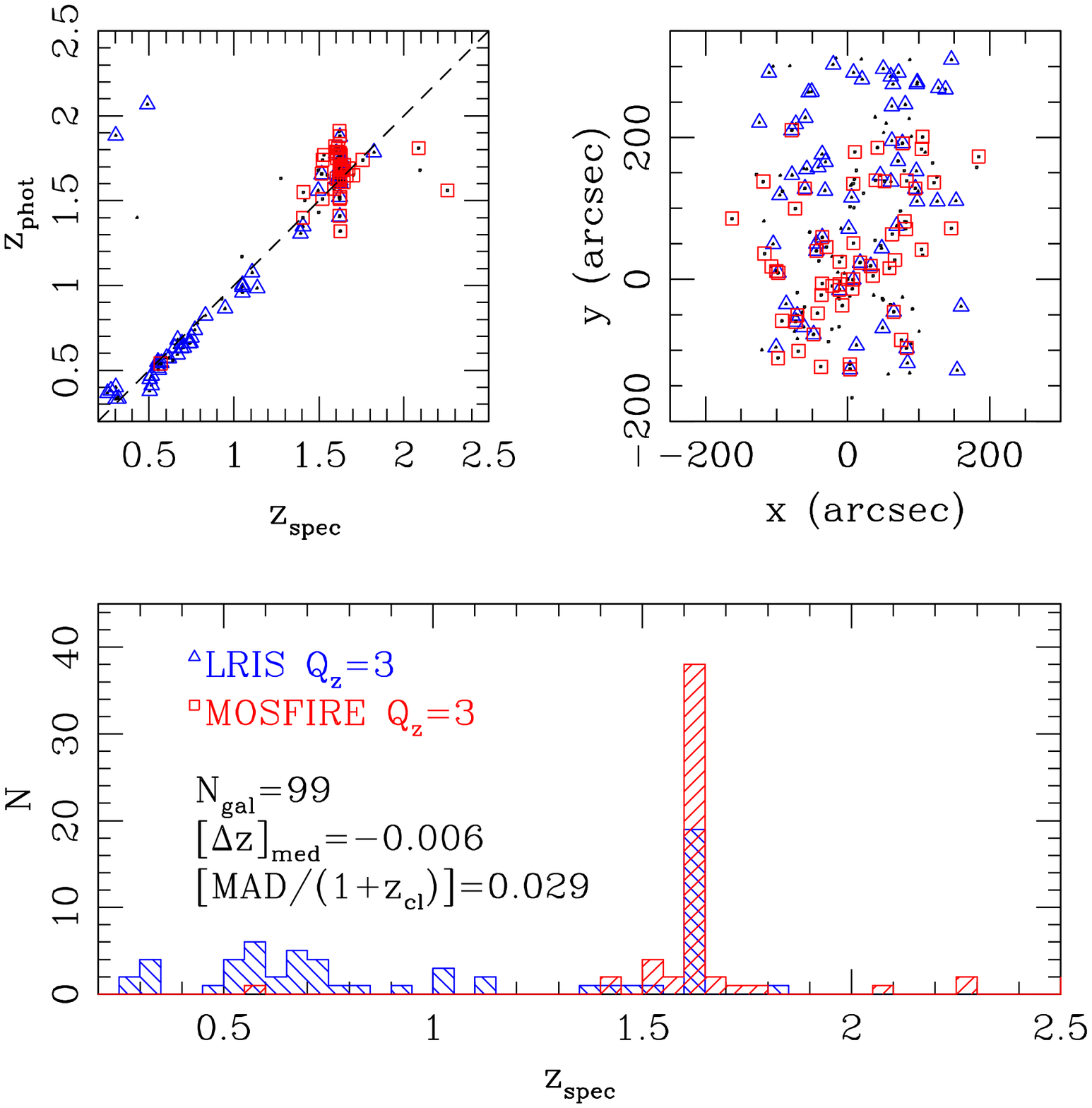}}
\caption{The photometric redshifts from \citet{williams:09}
compared to spectroscopic redshifts for all objects with measured
redshifts (top left; the dashed diagonal line denotes parity) and
their spatial distribution (top right).  Objects with \zspec\ quality
flag of $Q_z=3$ are shown as open blue triangles (LRIS) and open red
squares (MOSFIRE).  Note that $\sim30$ objects have \zspec\ from both
spectrographs.  The bottom panel shows the \zspec\ histograms for LRIS
(blue) and MOSFIRE (red) for objects with $Q_z=3$.
\label{fig:zhist}}
\end{figure*}

Each spectroscopic redshift is assigned a quality flag $Q_z$ where
$Q_z=3$ denotes a robust measurement (multiple spectral lines),
$Q_z=2$ is likely (single spectral line with potential secondary
line), and $Q_z=1$ is guess (single line and/or no strong spectral
features).  The combined spectroscopic observations from LRIS and
MOSFIRE yield 130 unique redshifts.  Note that all of the cluster
galaxies reported here are confirmed with emission-lines, e.g. the
split \OII$\lambda3727$\AA\ doublet or \Halpha+\NII\ pair.  Faint
continua are detected for only a handful of cluster galaxies and are
not reliable for measuring absorption-line redshifts, $i.e.$ our
survey is strongly biased towards active galaxies with emission lines.
In our analysis, we use only the 109 unique objects with redshift
quality flag of $Q_z=3$ (54 identified by LRIS and 55 by MOSFIRE).

We find the spectroscopic redshifts are in good agreement with the
photometric redshifts (Fig.~\ref{fig:zhist}, top left).  The
(\zphot-\zspec) distribution for the 99 objects with
\zspec$>0$ and Q$_{\rm z}=3$ has a median difference of $[\Delta {\rm
z}]_{\rm med}=-0.006$; we use the median to minimize the effect of
redshift outliers. The corresponding absolute median deviation is
$\sim3$\% where the latter is determined using $\Delta z/(1+z_{\rm
spec})$ and assuming \zspec=1.62.  For the 49 galaxies at
$1.5<$\zspec$<1.7$, the median difference is 0.06 and absolute median
deviation is also $\sim3$\%; there is only one spectroscopically
confirmed member with $|$\zphot-\zspec$|>0.3$.

The redshift distributions in Fig.~\ref{fig:zhist} (bottom)
demonstrate how well LRIS and MOSFIRE complement each other.  LRIS is
effective up to $z\sim1.6$ because \OII\ is still within the LRIS
wavelength range.  MOSFIRE is extremely effective at $z>1.5$ when
\OIII\ and \Halpha\ move into the J and H bands, respectively.

\subsection{NIR Emission-Line Fluxes}

To obtain spectro-photometric data and correct for the slit-loss, we
flux calibrate the MOSFIRE observations using a combination of
ground-based and \hubble\ imaging. We use the standard star flux
calibrated data from our custom IDL routines to compare with the data
from the (publicly) available UKIDSS Data Release 8 \citep{quadri:12}.
More information on the filter systems used in UKIDSS can be found in
\cite{hewett:06}.

We mask the sky regions in the spectra by assigning them weights of
zero and use an inverse variance weighting to calculate the total
spectroscopic fluxes for each 1D spectrum.  We use the UKIRT WFCAM
filter response functions to be able to directly compare our results
with the total broadband photometric fluxes of UKIDSS.

For each MOSFIRE mask in a given band, we calculate the median offset
between the spectroscopic magnitudes and the broadband photometric
magnitudes for all objects brighter than 23 mag. We use the median
offset as the correction factor to be applied to all the objects in
the respective masks to account for slit losses. We then calculate the
normalized median absolute deviation for objects with a broadband
magnitude brighter than 24 mag to determine the uncertainty for the
scaling process. We find the J-band to have systematically more
scatter compared to the H-band. We calculate the flux uncertainties in
the calibrated spectra to be $\sim$20\%.  A more detailed analysis
will be presented in Nanayakkara et al. (in prep).


\begin{figure*}
\centerline{\includegraphics[width=0.8\textwidth]{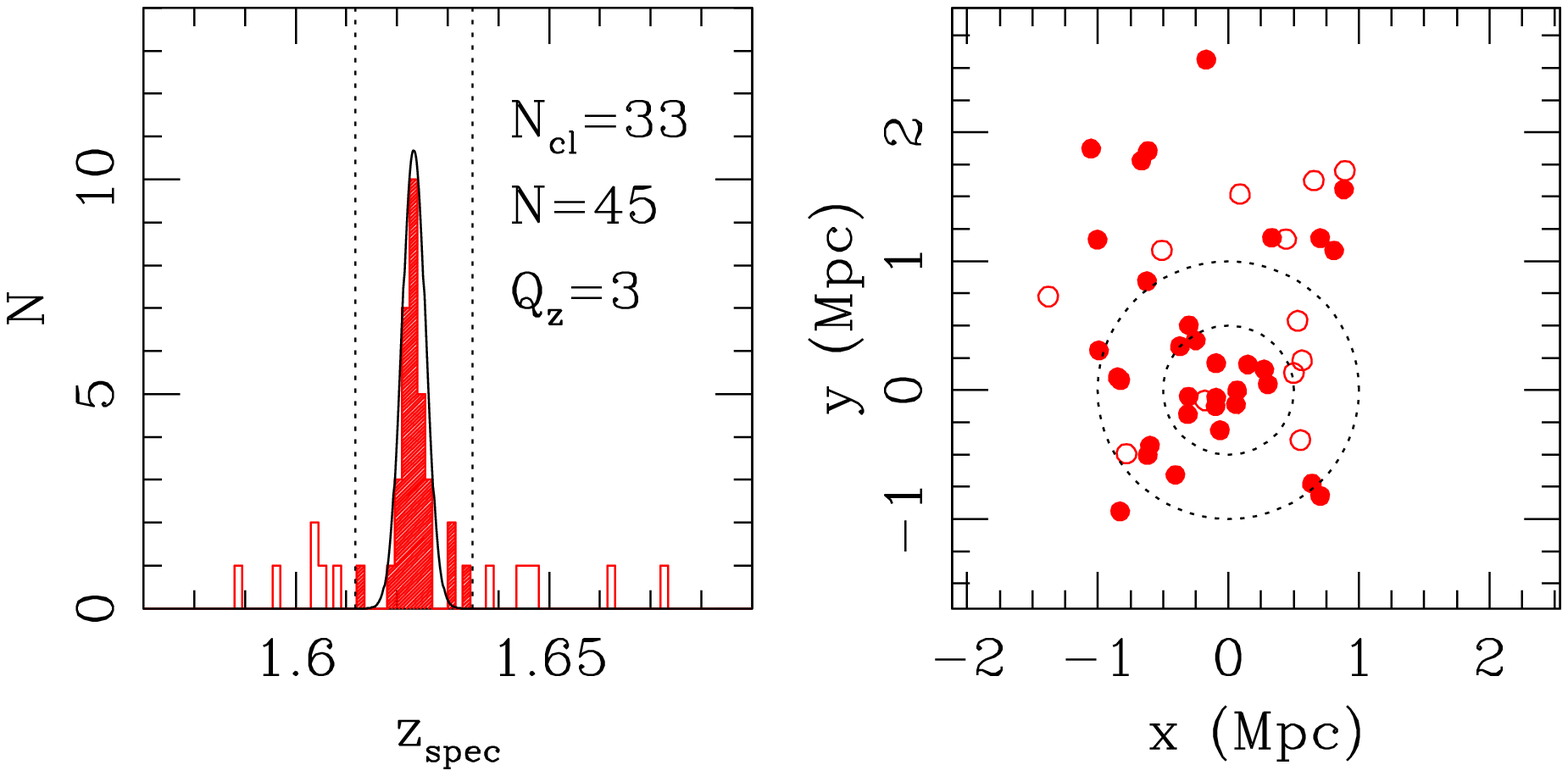}}
\caption{The galaxy cluster \irc's redshift range (left panel) is
determined by bootstrapping the 45 galaxies at $1.58<$\zspec$<1.68$.
We measure the cluster's median redshift and velocity dispersion to be
\zcl$\pm0.0003$ and \sigmacl$=254\pm50$~\kms; the latter is shown as a
Gaussian curve.  The cluster redshift range is defined by
$\pm3$\sigmacl\ (left panel, dotted vertical lines) which
isolates 33 emission-line members (filled histograms).  The
spatial distribution of the confirmed cluster members is shown in the
right panel (filled circles); galaxies outside the cluster redshift
range (open histogram) are shown as open circles.  At $z=1.62$, the
projected angular scale is $8.471$~kpc per arcsecond; the large dotted
circles correspond to projected (proper) radii at $z=1.62$ of 0.5 and
1 Mpc.
\label{fig:zhistcl}}
\end{figure*}

Our analysis uses \Halpha$\lambda6563$\AA, \Hbeta$\lambda4861$\AA,
\NII$\lambda6583$\AA, and \OIII$\lambda5007$\AA\ line fluxes to
measure redshifts, star formation rates, gas attenuation, and gas
metallicities of $z\sim1.6$ galaxies.  \Halpha\ is the strongest
emission line and falls in the H-band; all of the \Halpha\ detections
are measured at $>4\sigma$ significance for these galaxies.  We
identify 45 galaxies at $1.58<$\zspec$<1.68$ (Fig.~\ref{fig:zhistcl})
with \Halpha\ fluxes of $0.6-16.5$\ergu.  The H-band includes the much
weaker \NII\ emission line; objects with \NII\ measurements that are
below the detection limit of $\sim0.3$\ergu\ and/or detected with
statistical significance of $<2\sigma$ are assigned $1\sigma$ upper
limits from the line-fitting.

For our analysis of star formation and gas metallicities, we use the
narrow redshift range of $1.6118<$\zspec$<1.6348$ defined by our
kinematic analysis (\S\ref{sec:kinematics}) and exclude the cluster
galaxies that are X-ray detected (1) and that host broad-line AGN (1;
identified by \Halpha\ and \NII); this leaves 26 star-forming
cluster galaxies with \Halpha\ fluxes and a stellar mass range of
\logMstarMsun$=9.2-10.9$ (Table~\ref{tab:galcat}).  We consider
measurements for both individual galaxies as well as for stacked
spectra divided into three stellar mass bins.

\begin{figure*}
\centerline{\includegraphics[width=0.8\textwidth]{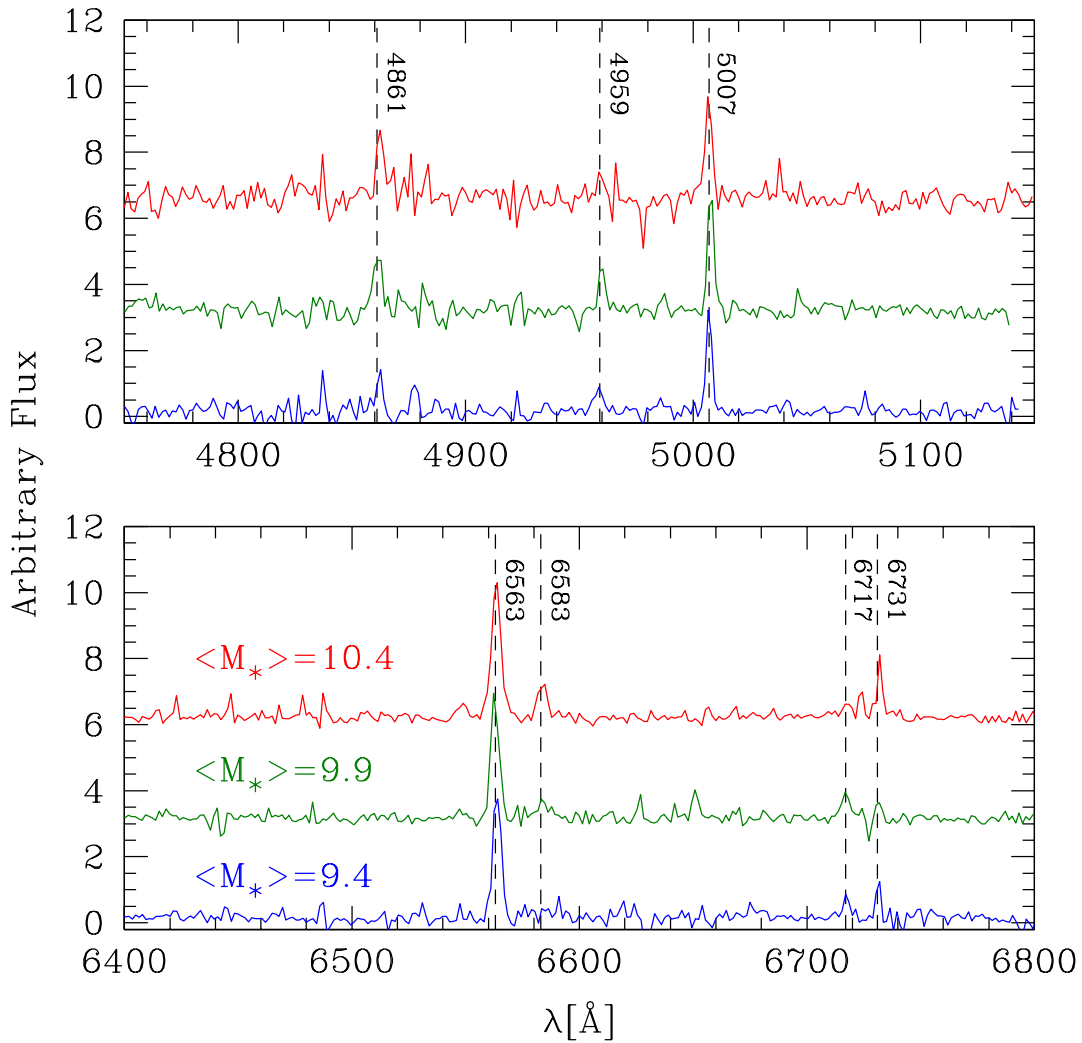}}
\caption{We divide the 26 \Halpha$\lambda6563$\AA-detected
cluster galaxies into three separate mass bins and stack their
spectra; for clarity, the stacked spectra are smoothed and offset in
flux.  The stellar mass range is \logMstarMsun$=9.2-10.9$
(Table~\ref{tab:SFR}) and the median stellar mass for each bin is
included in the figure; from low to high mass bin, there are 9, 9, and
8 members.  The individual spectra are first normalized by the
\Halpha\ flux and converted to rest-frame values, and then the median
is taken.  At \zcl, the spectral range corresponding to MOSFIRE
includes \Hbeta$\lambda4861$\AA\ and the
\OIII$\lambda\lambda4959,5007$\AA\ pair (top; J-band), and \Halpha, 
\NII$\lambda6583$\AA, and the  \SII$\lambda\lambda6717,6731$\AA pair
(bottom; H-band).
\label{fig:specstack}}
\end{figure*}

To stack the spectra, we first divide the 26 star-forming members into
three stellar mass bins with ranges of \logMstarMsun$=9.2-9.6$ (9
members), \logMstarMsun$=9.6-10.2$ (9 members), and
\logMstarMsun$=10.2-10.9$ (8 members).  We normalize the individual
galaxies by their \Halpha\ flux to give them equal weight
\citep[e.g.][]{sanders:15}, convert the spectra to rest-frame values,
combine the spectra, and then measure the line fluxes by fitting
Gaussian profiles.  Confidence intervals for the line fluxes are
determined by boot-strapping the 1D spectra.  The stacked spectra show
even the weaker emission lines \Hbeta\ and \NII\ as well as the \SII\
doublet (Fig.~\ref{fig:specstack}).

\begin{deluxetable*}{lrrrrrrrrrrr}
\tablecaption{Properties of \Halpha-detected \irc\
Members\tablenotemark{a}\label{tab:galcat}} 
\tablewidth{0pt}
\tablehead{
\colhead{Keck ID} 	& 
\colhead{$\alpha(2000)$} 	& \colhead{$\delta(2000)$} &
\colhead{$I$ mag} 	& \colhead {\zspec}	& 
\colhead{\logMstarMsun}	& \colhead {\ACalsed} & 
\colhead{f\Halpha\tablenotemark{b}}	&
\colhead{f\Hbeta\tablenotemark{b}}	&
\colhead {f\NII\tablenotemark{b}}	&
\colhead{SFR(\Halphased)\tablenotemark{c}}
}
\startdata
36395 & 34.61690 & -5.20299  & 24.70  & 1.613    & $ 9.41   ^{+ 0.18}_{-0.08} $ &  $ 0.2 ^{+0.3}_{- 0.2}$   & $ 1.8  \pm0.1 $ & $0.0  \pm0.6 $ & $  0.4  \pm0.3 $ &   2.2 \\
36849 & 34.56660 & -5.19900  & 25.00  & 1.624    & $ 9.60   ^{+ 0.07}_{-0.15} $ &  $ 0.1 ^{+0.3}_{- 0.1}$   & $ 3.4  \pm0.2 $ & $0.9  \pm0.4 $ & $  0.2  \pm0.2 $ &   3.6 \\
37269 & 34.56870 & -5.19590  & 24.72  & 1.624    & $10.39   ^{+ 0.05}_{-0.11} $ &  $ 0.1 ^{+0.5}_{- 0.1}$   & $ 5.1  \pm0.5 $ & $0.4  \pm0.5 $ & $  0.8  \pm0.3 $ &   5.3 \\
37522 & 34.60300 & -5.19369  & 24.95  & 1.622    & $10.20   ^{+ 0.17}_{-0.07} $ &  $ 1.1 ^{+0.2}_{- 0.5}$   & $ 2.1  \pm0.2 $ & $0.4  \pm0.3 $ & $  0.7  \pm0.1 $ &   8.1 \\
38080 & 34.61000 & -5.18868  & 24.15  & 1.625    & $10.34   ^{+ 0.00}_{-0.14} $ &  $ 0.1 ^{+0.2}_{- 0.1}$   & $10.3  \pm0.2 $ & $2.6  \pm0.5 $ & $  1.6  \pm0.2 $ &  11.0 \\
38455 & 34.60940 & -5.18625  & 24.02  & 1.623    & $10.87   ^{+ 0.02}_{-0.08} $ &  $ 0.8 ^{+0.1}_{- 0.2}$   & $ 4.6  \pm0.2 $ & $1.2  \pm0.3 $ & $  0.8  \pm0.2 $ &  12.2 \\
39150 & 34.59990 & -5.17829  & 25.50  & 1.624    & $ 9.22   ^{+ 0.14}_{-0.06} $ &  $ 0.2 ^{+0.3}_{- 0.2}$   & $ 2.3  \pm0.1 $ & $0.3  \pm0.5 $ & $  0.4  \pm0.1 $ &   2.8 \\
39463 & 34.59290 & -5.17625  & 24.23  & 1.622    & $ 9.57   ^{+ 0.05}_{-0.10} $ &  $ 0.1 ^{+0.1}_{- 0.1}$   & $ 8.7  \pm0.2 $ & $1.7  \pm0.6 $ & $  0.7  \pm0.1 $ &   9.2 \\
39771 & 34.59280 & -5.17402  & 24.36  & 1.623    & $ 9.81   ^{+ 0.07}_{-0.16} $ &  $ 0.6 ^{+0.2}_{- 0.2}$   & $ 4.9  \pm0.3 $ & $1.3  \pm0.4 $ & $  1.6  \pm0.2 $ &  10.1 \\
39989 & 34.58750 & -5.17219  & 25.28  & 1.623    & $10.29   ^{+ 0.10}_{-0.05} $ &  $ 0.2 ^{+0.8}_{- 0.2}$   & $ 3.6  \pm0.3 $ & $0.0  \pm0.3 $ & $  2.5  \pm0.3 $ &   4.3 \\
40243 & 34.61680 & -5.16958  & 24.78  & 1.622    & $ 9.61   ^{+ 0.07}_{-0.05} $ &  $ 0.2 ^{+0.2}_{- 0.2}$   & $ 2.4  \pm0.2 $ & $0.3  \pm0.5 $ & $  0.4  \pm0.1 $ &   2.8 \\
40382 & 34.61750 & -5.16884  & 25.18  & 1.623    & $10.05   ^{+ 0.08}_{-0.15} $ &  $ 1.4 ^{+0.2}_{- 0.2}$   & $ 5.3  \pm0.2 $ & $1.1  \pm0.3 $ & $  1.1  \pm0.2 $ &  30.8 \\
40568 & 34.58070 & -5.16690  & 24.79  & 1.626    & $ 9.49   ^{+ 0.08}_{-0.05} $ &  $ 0.0 ^{+0.3}_{- 0.0}$   & $ 6.7  \pm0.3 $ & $1.4  \pm0.2 $ & $  0.8  \pm0.1 $ &   6.2 \\
40729 & 34.58480 & -5.16558  & 24.76  & 1.627    & $ 9.49   ^{+ 0.01}_{-0.12} $ &  $ 0.1 ^{+0.2}_{- 0.1}$   & $ 2.0  \pm0.4 $ & $0.2  \pm0.4 $ & $  0.0  \pm0.1 $ &   2.1 \\
40731 & 34.59280 & -5.16529  & 24.84  & 1.625    & $ 9.25   ^{+ 0.04}_{-0.13} $ &  $ 0.0 ^{+0.3}_{- 0.0}$   & $ 2.7  \pm0.1 $ & $2.2  \pm0.7 $ & $  0.5  \pm0.2 $ &   2.5 \\
41189 & 34.62220 & -5.16206  & 23.77  & 1.623    & $ 9.78   ^{+ 0.09}_{-0.05} $ &  $ 0.2 ^{+0.2}_{- 0.2}$   & $ 9.0  \pm0.5 $ & $2.8  \pm0.3 $ & $  1.4  \pm0.2 $ &  10.9 \\
41297 & 34.60190 & -5.16097  & 24.44  & 1.622    & $ 9.93   ^{+ 0.11}_{-0.16} $ &  $ 0.8 ^{+0.2}_{- 0.2}$   & $ 5.3  \pm0.3 $ & $0.5  \pm0.4 $ & $  0.8  \pm0.1 $ &  14.0 \\
41548 & 34.59790 & -5.15948  & 24.48  & 1.624    & $10.50   ^{+ 0.15}_{-0.15} $ &  $ 1.0 ^{+0.3}_{- 0.3}$   & $ 7.3  \pm0.3 $ & $0.1  \pm0.5 $ & $  2.0  \pm0.3 $ &  25.1 \\
41956 & 34.59960 & -5.15566  & 23.75  & 1.621    & $10.25   ^{+ 0.05}_{-0.01} $ &  $ 1.0 ^{+0.1}_{- 0.2}$   & $14.6  \pm0.4 $ & $2.6  \pm0.8 $ & $  4.3  \pm0.2 $ &  50.2 \\
43178 & 34.61020 & -5.14441  & 24.66  & 1.621    & $10.47   ^{+ 0.13}_{-0.30} $ &  $ 1.2 ^{+0.6}_{- 0.3}$   & $ 5.4  \pm0.3 $ & $1.1  \pm0.5 $ & $  2.0  \pm0.2 $ &  24.3 \\
44102 & 34.56320 & -5.13659  & 24.62  & 1.621    & $10.69   ^{+ 0.20}_{-0.14} $ &  $ 0.4 ^{+1.4}_{- 0.2}$   & $13.8  \pm0.3 $ & $1.0  \pm0.6 $ & $  4.1  \pm0.2 $ &  21.5 \\
44403 & 34.57880 & -5.13336  & 24.88  & 1.622    & $ 9.36   ^{+ 0.11}_{-0.00} $ &  $ 0.2 ^{+0.2}_{- 0.2}$   & $ 4.9  \pm0.3 $ & $2.0  \pm0.6 $ & $  0.3  \pm0.1 $ &   5.9 \\
44459 & 34.62260 & -5.13383  & 23.93  & 1.624    & $10.12   ^{+ 0.12}_{-0.13} $ &  $ 0.4 ^{+0.3}_{- 0.2}$   & $ 7.0  \pm0.3 $ & $0.2  \pm0.5 $ & $  0.5  \pm0.3 $ &  11.0 \\
44587 & 34.56660 & -5.13347  & 24.12  & 1.621    & $ 9.92   ^{+ 0.03}_{-0.14} $ &  $ 0.5 ^{+0.2}_{- 0.1}$   & $ 4.9  \pm0.3 $ & $2.2  \pm0.5 $ & $  0.5  \pm0.1 $ &   8.8 \\
45959 & 34.56070 & -5.12105  & 25.05  & 1.620    & $ 9.28   ^{+ 0.22}_{-0.06} $ &  $ 1.6 ^{+0.0}_{- 0.3}$   & $ 2.0  \pm0.2 $ & $0.3  \pm0.4 $ & $  0.0  \pm0.1 $ &  15.2 \\
46922 & 34.61150 & -5.11372  & 23.21  & 1.630    & $10.16   ^{+ 0.14}_{-0.07} $ &  $ 0.4 ^{+0.3}_{- 0.1}$   & $14.9  \pm0.4 $ & $4.2  \pm0.4 $ & $  5.7  \pm0.3 $ &  23.6 \\
\enddata
\tablenotetext{a}{\irc\ members ($1.6118<$\zspec$<1.6348$;
see \S\ref{sec:kinematics}) that have \Halpha\ emission as measured
with MOSFIRE.}
\tablenotetext{b}{Observed fluxes in units of
$10^{-17}$~erg~s$^{-1}$~cm$^{-2}$.  Note that in several cases,
f\Hbeta\ and f\NII\ are upper limits considering the S/N.}
\tablenotetext{c}{\Halpha\ fluxes are corrected for dust using \Ased\
(Eq.~\ref{eq:AHa}; see \S\ref{sec:balmer}) and then converted to star
formation rates using the relation from
\citet{hao:11}.} 
\end{deluxetable*}

\subsection{Stellar Masses \& Stellar Attenuation}

To determine stellar masses (\Mstar) and stellar attenuation
(\ACalsed; Table~\ref{tab:galcat}), we use the catalog of the UDS
field from \citet{quadri:12} with our spectroscopic redshifts.  Both
\Mstar\ and \ACalsed\ are determined by using FAST \citep{kriek:09a}
to fit Spectral Energy Distributions (SEDs) to the multi-wavelength
imaging.  We use a \citet{chabrier:03} initial stellar mass function,
constant solar metallicity, and exponentially declining star formation
rate ($\tau=$10 Myr to 10 Gyr).  

We also use FAST to determine the
corresponding errors for \Mstar\ and
\ACalsed; the latter assumes \Rv=4.05 \citep[starburst attenuation
curve;][]{calzetti:00}\footnote{The starburst (SB) attenuation curve
is commonly referred to as the Calzetti law and is appropriate for
continuum measurements.  We use ``starburst'' as requested by
D. Calzetti.}.  FAST determines confidence intervals by using
Monte Carlo simulations and modifying the observed fluxes by their
photometric errors in each simulation.  Note that there are
degeneracies between stellar attenuation, star formation histories,
metallicities, and dust laws calculated by FAST.  These degeneracies
can sometimes be broken by independently measuring, e.g. nebular
attenuation.  For a subsample of the galaxies, we measure both
\Halpha\ and \Hbeta\ emission to determine the Balmer decrement and
thus compare nebular to stellar attenuation (\S\ref{sec:balmer}).


\begin{figure*}
\centerline{\includegraphics[width=0.8\textwidth]{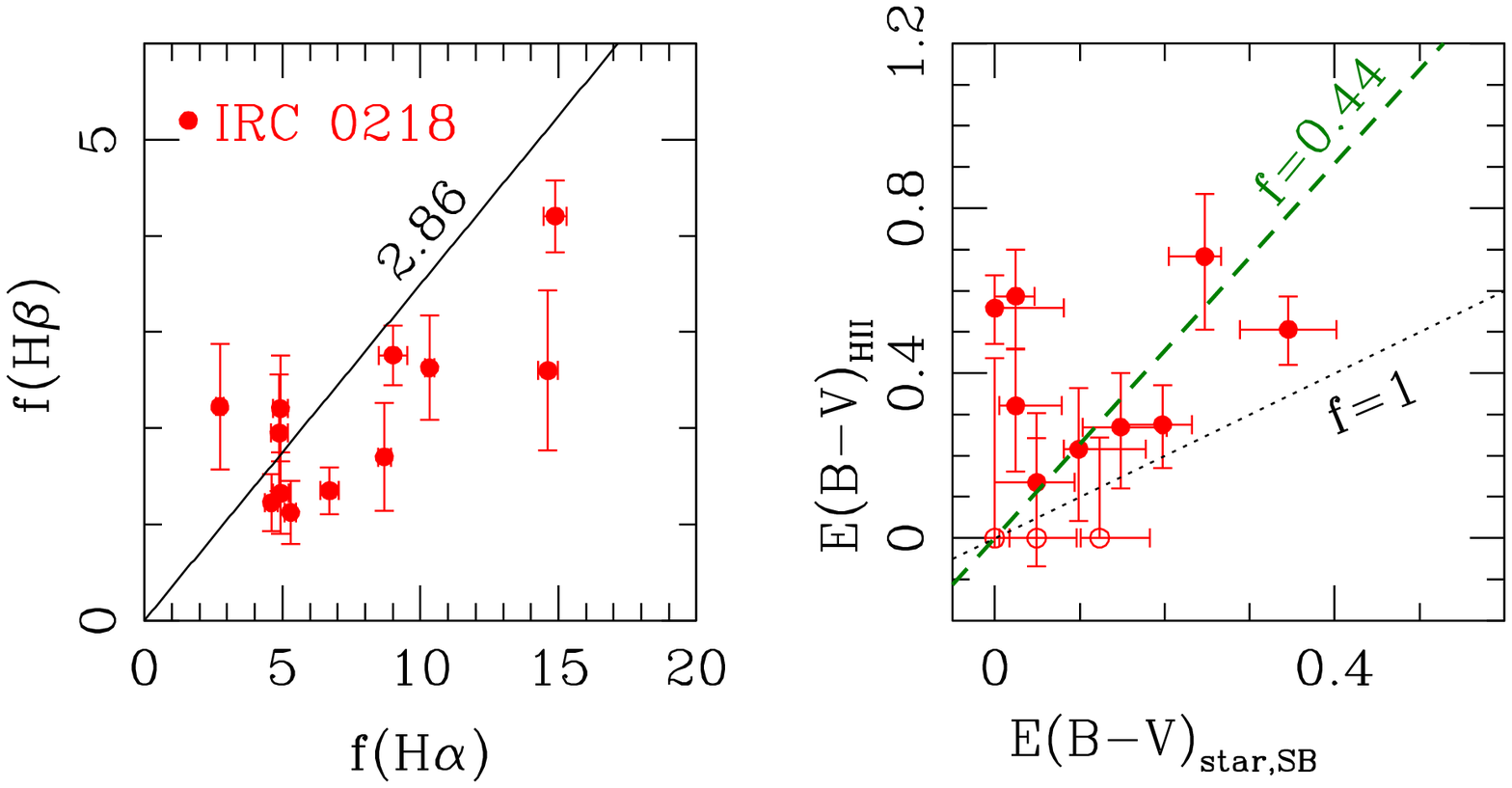}}
\caption{{\it Left:} Distribution of \Halpha\ vs. \Hbeta\ line fluxes
(\ergu) as measured with MOSFIRE where the diagonal line corresponds
to \Halpha/\Hbeta$=2.86$ (Balmer decrement for Case B recombination).
Our requirement of \Hbeta$\geq0.5$\ergu\ measured at $\geq3\sigma$
confidence identifies 12 cluster galaxies (filled circles); included
are the $1\sigma$ errors determined from propagating the error
spectra.  
The attenuation for the the 3 cluster galaxies with
\Halpha/\Hbeta$<2.86$ is likely negligible and so we set their Balmer
decrement to 0. 
{\it Right:} The color excess measured using the Balmer
decrement compared to the color excess from the stellar continuum
measured with FAST; we include only the 9 cluster galaxies with
\Halpha/\Hbeta$\geq2.86$ and use the CCM89 reddening law to calculate
the nebular \EBV.  For reference, we include the ratio of stellar to
nebular \EBV\ measured for local starburst galaxies where the color
excess for stars is less \citep[$f=0.44$, dashed
line;][]{calzetti:00}.  The cluster galaxies at $z\sim1.6$ follow this
general trend, but there is signicant scatter that mirrors recent
results based on larger samples of field galaxies
\cite[see][]{reddy:15}.
\label{fig:ebv}}
\end{figure*}

\section{Analysis}

\subsection{Galaxy Cluster Kinematics}\label{sec:kinematics}

Our combined LRIS and MOSFIRE spectroscopic survey confirms 109 unique
objects with redshift quality flag $Q_z=3$; of these, nine are stars.
We identify 45 galaxies at $1.58<$\zspec$<1.68$ and the galaxy cluster
is a well-defined peak in the redshift distribution at $z\sim1.6$
(Fig.~\ref{fig:zhist}, bottom).  To determine the cluster
redshift and redshift range, we first need to identify the members.
We remove outliers iteratively using a median absolute deviation
(MAD); the MAD is less sensitive to outliers that can be problematic
for dynamically young systems such as \irc.  For a gaussian
distribution, $\sigma\approx1.48\times$MAD and we use $3\sigma$ limits
to define the redshift range.  We employ 5 iterations and find a
stable solution requires only 2 iterations: \irc\ has 33 members
within $1.6118<$\zspec$<1.6348$ which corresponds to limits defined by
$\pm3\sigma$ (Fig.~\ref{fig:zhistcl}).

To determine \irc's redshift, we bootstrap the 33 galaxies
\citep[15000 realizations;][]{beers:90} and measure a median cluster
redshift of \zcl$\pm0.0003$ and mean cluster redshift of
$1.6232\pm0.0004$. The cluster galaxies have a strikingly narrow
redshift distribution and the cluster velocity dispersion (from the
bootstrapped distribution) is correspondingly low:
\sigmacl$=254\pm50$~\kms, $i.e.$ more like a group than a fully
developed galaxy cluster (see \S\ref{sec:dispersion}).
Figure~\ref{fig:zhistcl} (right) shows that most (20/33) of the
cluster members are within \Rproj$<1$~Mpc of the BCG.  If we consider
only the members at Rproj$<0.5$~Mpc, the cluster redshift and velocity
dispersion are virtually identical.  The combination of high spatial
density and low velocity dispersion are very conducive to
galaxy-galaxy merging \citep{rudnick:12,lotz:13}, $i.e.$
frequency and duration of galaxy-galaxy interations is higher than in
a field environment. 

\irc's redshift and velocity dispersion do not change
significantly if we limit our analysis to either the MOSFIRE or LRIS
observations only.  If we repeat our analysis using only the MOSFIRE
redshifts, there are 30 members: the median cluster redshift remain
the same and the velocity dispersion decreases slightly to
$\sigma_{\rm cl,MOS}=220\pm43$~\kms.  Using only the LRIS
redshifts, there are 17 members: the median cluster redshift remains
the same and the velocity dispersion increases slightly to
$\sigma_{\rm cl,LRIS}=305\pm118$~\kms.

\subsection{Balmer Decrement and Attenuation}\label{sec:balmer}

For a subset of MOSFIRE-detected galaxies, we are able to measure both
\Hbeta\ and \Halpha\ and thus directly determine their internal
extinction via the Balmer decrement.  \Halpha\ is by far the stronger
line (Fig.~\ref{fig:specstack}) where \Halpha/\Hbeta=2.86 is the
intrinsic line flux ratio for Case B recombination with electron
temperature $T_e=10^4$~K and electron density of $n_e=100$~cm$^{-3}$
\citep{osterbrock:89}.  In addition to being weaker, at $z\sim1.6$ the
\Hbeta\ line falls in a region of telluric absorption ($\lambda_{\rm
obs}\sim1.264\mu$m) which makes it even more challenging to measure.

Only 12 cluster galaxies meet our selection criteria of
\Hbeta$\geq0.5$\ergu\ measured at $\geq3\sigma$ significance
(Fig.~\ref{fig:ebv}, left). There are 9 cluster galaxies with
\Halpha/\Hbeta$\geq2.86$ and three members with lesser values that are
within $\sim1-2\sigma$ of this ratio (Fig.~\ref{fig:ebv}, left).  The
latter are likely to have negligible attenuation and so we set their
attenuation to zero.  

The total attenuation curve is parameterized by

\begin{equation}
{\rm R}_{\rm V} \equiv \frac{{\rm A}_{\rm V}}{{\rm E(B-V)}}
\end{equation}

\begin{equation}
k(\lambda) = \frac{{\rm A}(\lambda)}{{\rm E(B-V)}}
\end{equation}

where A$(\lambda)$ is the total magnitude of the extinction at
wavelength $\lambda$ and \EBV\ is the color excess as measured by
nebular lines or the stellar continuum.  For galaxies with a measured
Balmer decrement, the color excess is determined directly from the
observed line fluxes with

\begin{equation}
{\rm E(B-V)_{\rm HII}} = 
\left[k({\rm H}\beta)-k({\rm H}\alpha)\right]^{-1}
2.5 \log_{10}\left( \frac{{\rm H}\alpha/{\rm H}\beta}{2.86}\right)
\end{equation}

Using the \citet[CCM;][]{cardelli:89} attenuation curve for the
diffuse interstellar medium where R$_{\rm V}=3.1$, $k({\rm
H}\alpha)=2.53$, and $k({\rm H}\beta)=3.61$, the color excess is then

\begin{equation}
{\rm E(B-V)_{\rm HII}} = 2.33 
\log_{10}\left( \frac{{\rm H}\alpha/{\rm H}\beta}{2.86}\right).
\end{equation}

Figure~\ref{fig:ebv} (right) compares the color excess of the ionized
gas \EBVgas\ to that of the stellar population \EBVsed; the latter is
determined by fitting to the (UV) stellar continuum with FAST (solar
metallicity) and assuming \CalRv=$4.05$ \citep{calzetti:00}.  For the
error in \EBVgas, we add the \Halpha\ and \Hbeta\ line flux errors in
quadrature.  For the cluster galaxies, the nebular \EBVgas\ is larger
than that of the stellar component \EBVCalsed.  This is consistent
with results at $z\sim0$ and $z\sim1.4$ that support a ratio of
\EBVsed\ to \EBVgas\ of 0.44 \citep{calzetti:00,price:14}. 
However, we note the large scatter in this ratio that mirrors results
from recent studies of field galaxies at $z\sim2$ \citep{reddy:15}.
Although it is tempting to measure this ratio using the cluster
galaxies, our limited sample does not allow us to confirm nor exclude
the ratio of 0.44.


\begin{figure*}
\centerline{\includegraphics[width=0.8\textwidth]{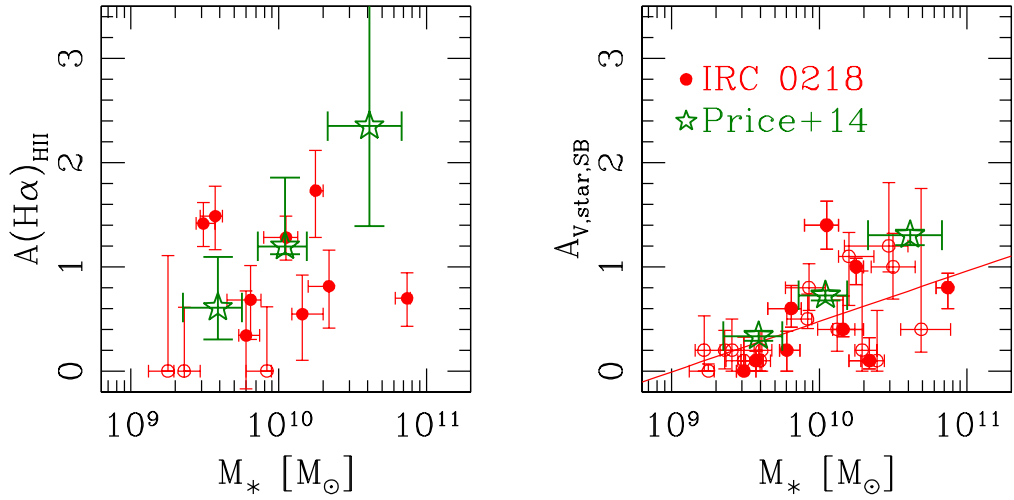}}
\caption{The nebular (left) and stellar (right) attenuation versus
stellar mass for individual cluster galaxies with measured Balmer
decrement (filled circles) and \ACalsed\ only (open circles).  Shown
for comparison are the stacked values for the nebular (left) and
stellar (right) attenuation of the field at $z\sim1.4$ \citep[open
stars;][]{price:14}.  The \irc\ galaxies do not show any trend between
nebular attenuation \Agas\ and stellar mass; this is also true when
stacking the cluster spectra.  In contrast, the cluster galaxies do
increase in stellar attenuation \ACalsed\ with \Mstar\ as confirmed
with a least-squares fit (diagonal line), but the cluster has a
shallower slope than in the field \cite[see also][]{reddy:15}.
\label{fig:Av_smass}}
\end{figure*}

Figure~\ref{fig:Av_smass} (left) shows the cluster galaxies that
satisfy our rigorous \Hbeta\ criteria and compares their nebular
attenuation to stellar mass.  The members do not define a correlation
between \Agas\ and \Mstar, in contrast to results at $z\sim0.1$
\citep{garn:10} and $z\sim1.4$ \citep{price:14}, nor is any trend
visible when we stack the spectra into three mass bins.

The cluster galaxies define a clearer trend between stellar
attenuation \ACalsed\ as measured by FAST (Fig.~\ref{fig:Av_smass},
right) and \Mstar.  The least-squares fit to the 20
members\footnote{We exclude X-ray and broad-line AGN from the fit and
also member 45959 whose SED fit indicates a very dusty low-mass galaxy
(see Table~\ref{tab:galcat}).}  is

\begin{equation}
{\rm A}_{\rm V,star,SB}=0.48\log[{\rm M}_{\ast}/{\rm
M}_{\odot}] - 4.36
\end{equation}

where $\sigma_{\rm RMS}=0.34$.  A fit using only the 9 members with a
measured Balmer decrement is identical.  The relation between
\ACalsed\ and \Mstar\ seems to be shallower than measured for the
field at $z\sim1.4$ \citep{price:14}, but there is large scatter.

How then can we correct our \Halpha\ fluxes for attenuation given that
for most of the \Halpha-detected galaxies, we only have the stellar
attenuation \ACalsed\ measured by FAST?  For simplicity,
we first correct the \Halpha\ line fluxes using the nebular attenuation
curve from \citet{cardelli:89} with \Rv=3.1 

\begin{equation}
{\rm A}({\rm H}\alpha)_{\rm HII} = 2.53 \times {\rm E(B-V)}_{\rm HII}.
\end{equation}

We then use the observed stellar to nebular attenuation ratio of
\EBVsed$=0.44\times$\EBVgas\ \citep{calzetti:00} and the stellar
attenuation \ACalsed\ as measured by FAST for all of the galaxies
where \EBVsed=\ACalsed/4.05 \citep[see also][]{steidel:14}

\begin{equation}
{\rm A}({\rm H}\alpha)_{\rm HII} = 5.75 \times {\rm E(B-V)}_{\rm
star}\label{eq:AHa}
\end{equation}

We use Eq.~\ref{eq:AHa} to correct all of the \Halpha\ fluxes for
attenuation and determine corresponding star formation rates.
Recently \citet{reddy:15} measure \Rv=2.505 for field galaxies at
$z\sim2$, a value that is lower than the \Rv\ for the starburst and
diffuse ISM attenuation curves.  However, we cannot discriminate
between these models given our limited cluster sample.

\subsection{\Halpha\ Star Formation Rates}

To determine \Halpha\ star formation rates, we correct for attenuation
using \EBVsed\ (Eq.~\ref{eq:AHa}) and convert the \Halpha\ fluxes to
star formation rates using the relation from \citet{hao:11}:

\begin{equation}
\log[{\rm SFR}({\rm H}\alpha_{\rm star})] = \log[{\rm L}({\rm
H}\alpha_{\rm star})] - 41.27
\end{equation}

This relation assumes a Kroupa IMF
\citep[$0.1-100$\Msun;][]{kroupa:01}, but the relation for a Chabrier
IMF is virtually identical (difference of 0.05).  The uncorrected
\Halpha\ fluxes of the 26 star-forming members range from
$1.8-14.9$\ergu.  Using \EBVsed\ to correct for attenuation, the
corresponding \Halpha\ star formation rates are $2-50$\Msunyr\
(Table~\ref{tab:galcat}).  We have excluded X-ray detected members 
and broad-line AGN. 

The cluster field was imaged with Spitzer MIPS as part of the
legacy UKIDSS Ultra Deep Survey (SpUDS PI: J.~Dunlop) and the catalog
includes all sources detected with $S/N>5$ which corresponds to a flux
of $\sim40\mu$Jy. The \mipsmu\ fluxes are converted to total infrared
luminosities (L$_{IR}$) using the \citet{chary:01} templates, and SFRs
are calculated from $L_{IR}$ using the prescription of
\citet{kennicutt:98}, adjusted to the Chabrier IMF.  We are able to
measure \mipsmu\ fluxes for the six cluster members with the highest
\Halpha\ ($>20$\Msunyr).

Even with correcting for dust attenuation, the \Halphased\ SFRs
are lower compared to the \mipsmu\ SFRs: on average, the \mipsmu\ SFRs
for the six detected members ($42-155$\Msunyr) are a factor of $\sim3$
larger.  The advantage of using \Halpha\ is that we are significantly
more sensitive, i.e. we can measure SFRs to $\sim2$\Msunyr\ compared
to $\sim40$\Msunyr\ with the \mipsmu\ observations, and thus we are
not limited to only the most strongly star-forming dust-obscured
cluster galaxies.  \citet{papovich:07} also showed that for galaxies
at $z\sim2$, SFRs based on \citet{chary:01} templates fit only to
\mipsmu\ are larger by $\sim2-5$ compared to SFRs based on a
combination \mipsmu, $70\mu$m, and $160\mu$m.  This is intriguingly
consistent with the offset we measure when comparing our \Halphased\
SFRs to the \mipsmu\ SFRs.

\begin{figure*}
\centerline{\includegraphics[width=0.8\textwidth]{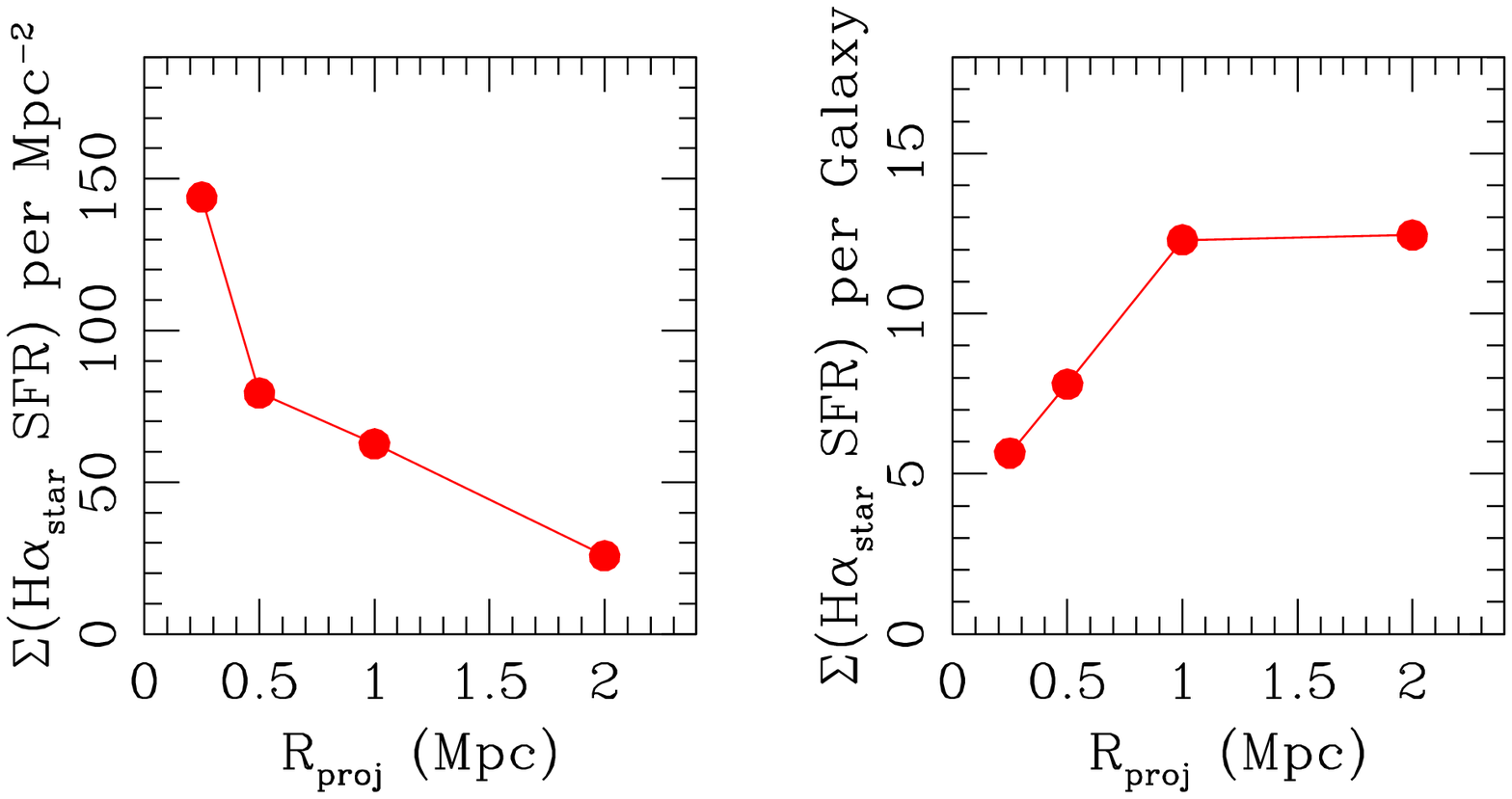}}
\caption{The integrated \Halphased\ star formation rate per unit area
(the total\Msunyr\ per square Mpc) is highest in the cluster core
(\Rproj$<0.25$~Mpc; left) and decreases with increasing cluster radius
\citep[e.g.][]{tran:10,brodwin:13}.  This is driven by the existence
of star-forming members in the core, but the {\it average} \Halphased\
SFR per galaxy is actually lowest in the cluster core and plateaus at
\Rproj$\gtrsim1$~Mpc (right; Table~\ref{tab:SFR}).  The propagated
errors due to the uncertainty in the \Halpha\ line fluxes are smaller
than the points ($\pm0.6-1.4$\Msunyr).
\label{fig:ISFR}}
\end{figure*}

\begin{figure*}
\centerline{\includegraphics[width=0.8\textwidth]{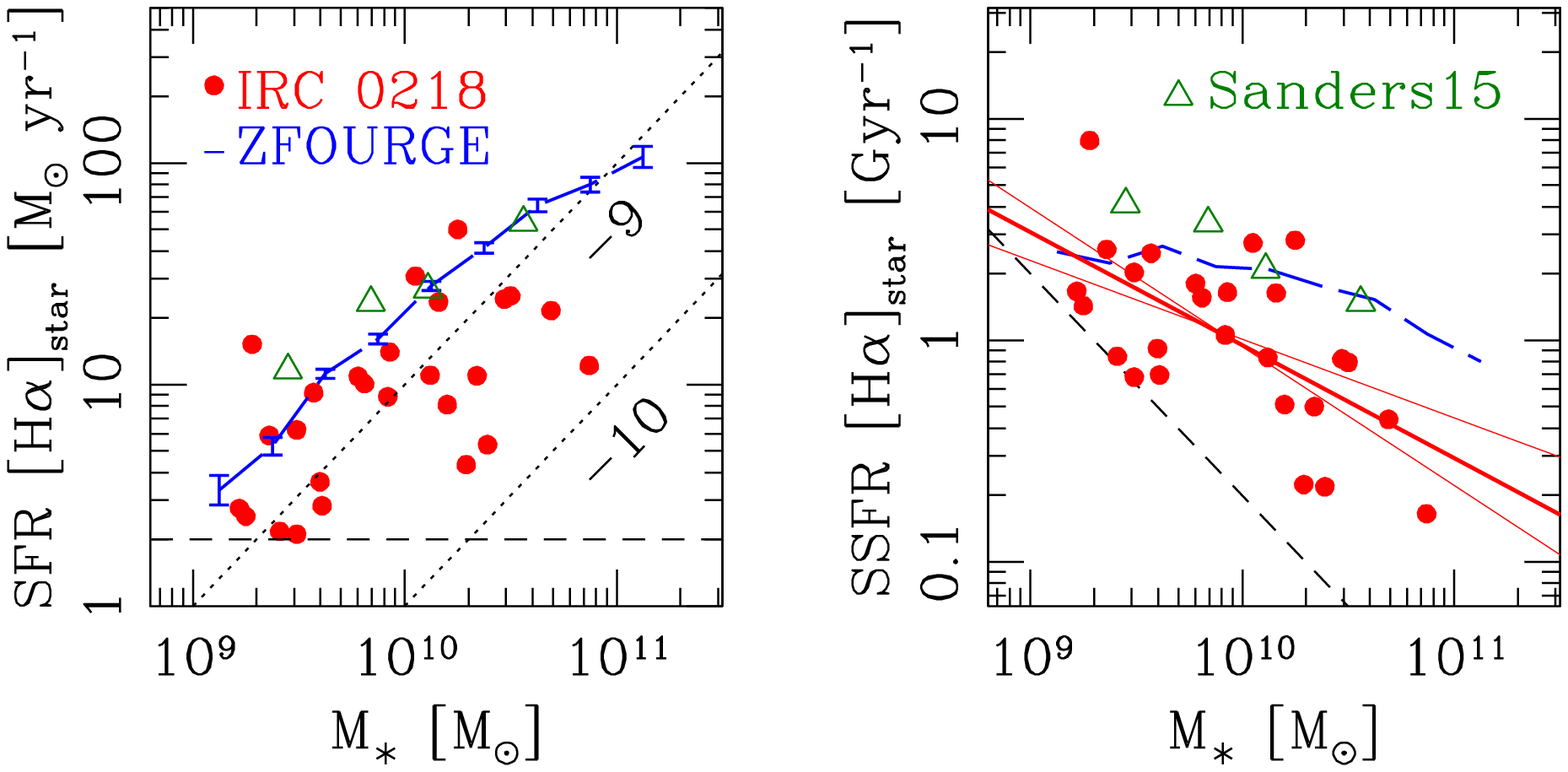}}
\caption{The 26 cluster galaxies tend to have lower \Halphased\
star formation rates (left) and specific SFRs (right) at a given
stellar mass compared to their field counterparts.  {\it Left:} The
\Halphased\ SFRs for the cluster galaxies (filled circles) are
corrected for extinction using \EBVsed\ and typical errors in
\Halphased\ SFRs and \Mstar\ are $\sim0.1-0.2$~dex (see
\S~\ref{sec:balmer}; Table~\ref{tab:galcat}). Included are lines of
constant specific star formation rates (SSFR; \Msunyr\ per \Mstar;
dotted lines) and the horizontal dashed line is our lower SFR limit.
For comparison, we include the binned points measured by MOSDEF for
\Halpha-detected field galaxies at $z\sim2.3$
\citep[open triangles;][]{sanders:15} and the star-forming field
galaxies at $1.5<z<2.0$ from ZFOURGE (long-dashed curve; Tomczak et
al, submitted; UV+MIR SFRs).  The ZFOURGE errors represent uncertainty
in the mean and the stellar mass bins are 0.25 dex.  {\it Right:} The
higher mass cluster galaxies tend to have lower SSFRs compared to
their field counterparts in both ZFOURGE (long-dashed curve) and
MOSDEF; a least-squares fit to the \irc\ members with SFRs$>2$\Msunyr\
(solid line; $1\sigma$ range shown as thin lines) measures a slope
that is steeper than in ZFOURGE.  The dashed diagonal line shows
the SSFR limit corresponding to our SFR limit of 2\Msunyr.
\label{fig:sfr_smass}}
\end{figure*}

We confirm the original measurement by \citet{tran:10} of an elevated
star formation rate within the cluster core: the integrated
\Halphased\ SFR per square Mpc at \Rproj$<0.25$~Mpc is about six times
higher than at \Rproj$<2$~Mpc (Fig.~\ref{fig:ISFR};
Table~\ref{tab:SFR}).  The five star-forming members at
\Rproj$<0.25$~Mpc have stellar masses of \logMstarMsun$=9.3-10.3$.
The cluster galaxies also show increasing SFR with stellar mass
(Fig.~\ref{fig:sfr_smass}).  We stress that the integrated \Halphased\
SFRs are lower limits because 1) for the six members with both
\Halphased\ and \mipsmu\ SFRs, the \mipsmu\ values are on average
$\sim3$ times higher; and 2) the limited spatial coverage of the NIR
spectroscopy combined with the increasing contamination by field
galaxies means we identify fewer \Halpha-emitting members at larger
\Rproj.  In \S\ref{sec:sfr_smass}, we compare in detail the cluster
\Halphased\ SFRs to that measured for the field at $z\sim2$.

\begin{deluxetable*}{lrrrrr}
\tablecaption{\Halphased\ Star Formation Rates vs. Cluster Radius\label{tab:SFR}}
\tablewidth{0pt}
\tablehead{
\colhead{\Rproj} & \colhead{N$_{\rm gal}$} & 
\colhead{$\Sigma$(\Halpha\ SFR)\tablenotemark{a}} &
\colhead{$\Sigma$(\Halpha\ SFR) per galaxy \tablenotemark{b}} &
\colhead{$<$SSFR$>$ \tablenotemark{c}} &
\colhead{$\Sigma$(\Halpha\ SFR) per Area }
}
\startdata
250 kpc & 5     & 28\Msunyr     &  5.6\Msunyr & 0.8 Gyr$^{-1}$ & 144 \Msunyr~Mpc$^{-2}$\\
500 kpc & 8     & 62\Msunyr     &  7.8\Msunyr & 0.9 Gyr$^{-1}$ & 79  \Msunyr~Mpc$^{-2}$\\ 
1 Mpc& 16       & 197\Msunyr    & 12.3\Msunyr & 0.8 Gyr$^{-1}$ & 63  \Msunyr~Mpc$^{-2}$\\ 
$<2$ Mpc& 26 	& 324\Msunyr    & 12.5\Msunyr & 0.9 Gyr$^{-1}$ & 26 \Msunyr~Mpc$^{-2}$\\ 
\enddata
\tablenotetext{a}{The total enclosed \Halpha\ star formation rate
corrected for attenuation using \EBVsed\ (see \S\ref{sec:balmer}). 
X-ray detected members and broad-line AGN are excluded.  Given
spectroscopic incompleteness at larger radii, we consider these to be
lower limits on \irc's total SFR.}
\tablenotetext{b}{The integrated \Halphased\ SFR divided by the number
of \Halpha-detected galaxies within \Rproj; quiescent members are
excluded.} 
\tablenotetext{c}{The specific star formation rate determined by
summing the \Halphased\ SFR and \Mstar\ for the \Halpha-detected
members within \Rproj; quiescent members are excluded.}
\end{deluxetable*}

\subsection{Gas-Phase Metallicities}

Following several recent analyses on gas metallicities in the field
\citep{kulas:13,steidel:14,sanders:15}, we measure the oxygen
abundance for the cluster galaxies using the N2 method calibrated by
\citet{pettini:04}: 

\begin{equation}
12 + \log {\rm (O/H)}_{\rm N2} = 8.90 + 0.57\times\log\left(
{\hbox{[{\rm N}\kern 0.1em{\sc ii}]}}/{\rm H}\alpha \right)
\end{equation}

We use only N2 to measure gas metallicities due to our limited ability
to measure \Hbeta\ and \OIII\ for individual galaxies.  We exclude
X-ray detected members and broad-line AGN; we also follow 
\citet{sanders:15} and require that $\log$(\NII/\Halpha)$<-0.3$ to
remove potential AGN which removes one member.  From our sample of
\Halpha-detected cluster galaxies, we measure N2 for the 20 members
where the \NII\ emission line is detected at (S/N)$>2$; note that
\NII\ is much weaker than \Halpha.  For five additional members, we
place $1\sigma$ upper limits on N2.  We also measure gas metallicities
for the binned spectra (Fig.~\ref{fig:specstack}).

\begin{figure*}
\centerline{\includegraphics[width=0.8\textwidth]{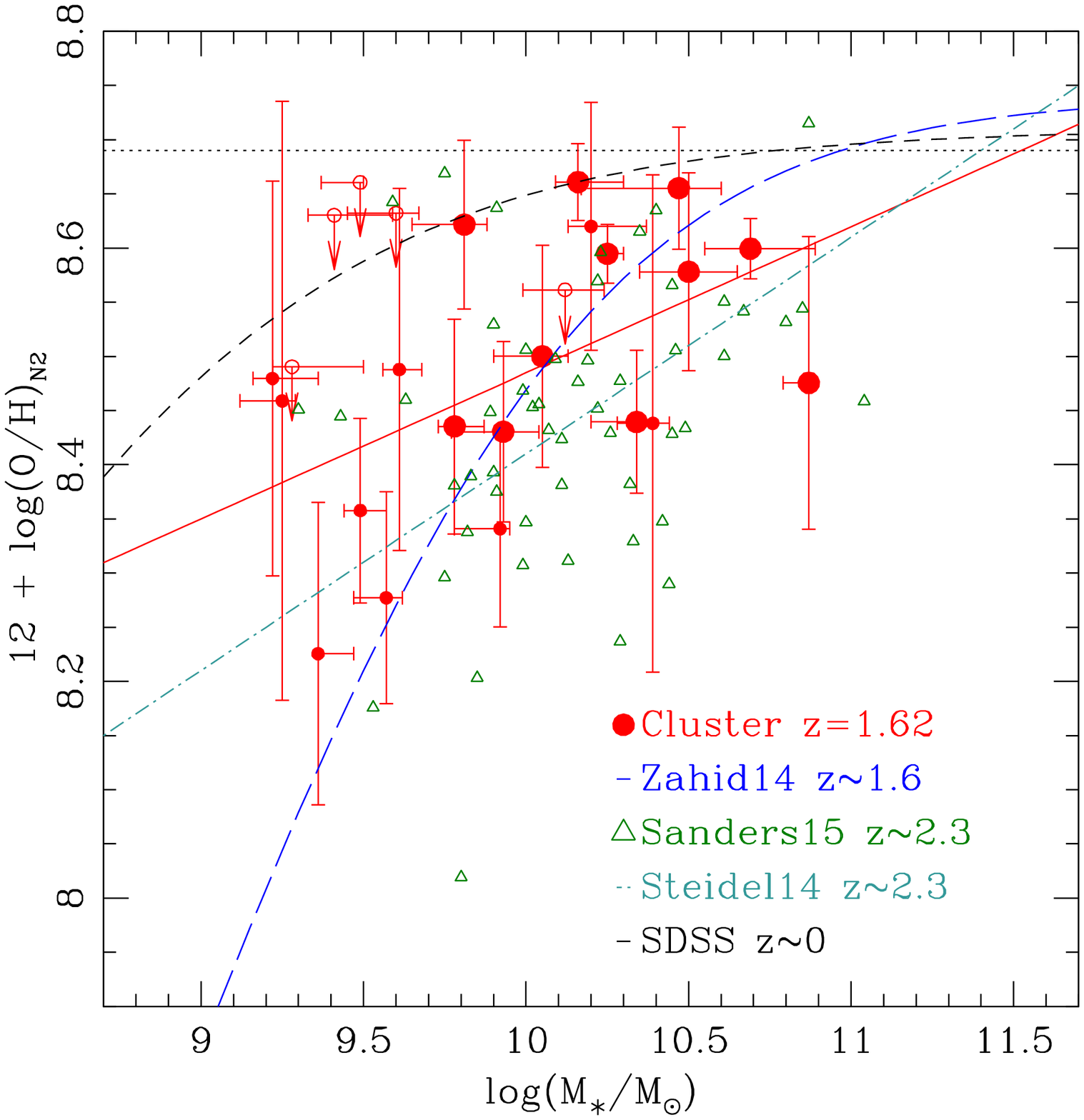}}
\caption{Stellar mass versus gas metallicity for \irc\  members
as measured by \NII/\Halpha\ (filled red circles); we fit a
least-squares to these galaxies (red line).  These cluster galaxies
are also divided by SFR[\Halphased]=10\Msunyr\ into low (small filled
red circles; 9) and high (large filled red circles; 11) star formation
rates.  Red open circles (5) with arrows are members with $1\sigma$
upper limits on \NII. Included for comparison are: the MZR of field
galaxies at $z\sim1.6$ in COSMOS \citep[blue long-dashed
curve;][]{zahid:14}; field galaxies at $z\sim2.3$ from MOSDEF
\citep[open green triangles;][]{sanders:15}; the MZR relation from
SDSS \citep[dashed black curve;][]{moustakas:11}; the linear fit to
$z\sim2.3$ galaxies from \citet[dot-dashed cyan line;][]{steidel:14};
and solar abundance \citep[dotted line at 8.69;][]{asplund:09}.  The
MZR relation for the cluster \irc\ (red line) is clearly offset from
local galaxies (dashed curve) and agrees with the Steidel fit to
galaxies at $z\sim2.3$ (dot-dashed cyan line).
\label{fig:mzr}}
\end{figure*}

\begin{figure*}
\centerline{\includegraphics[width=0.8\textwidth]{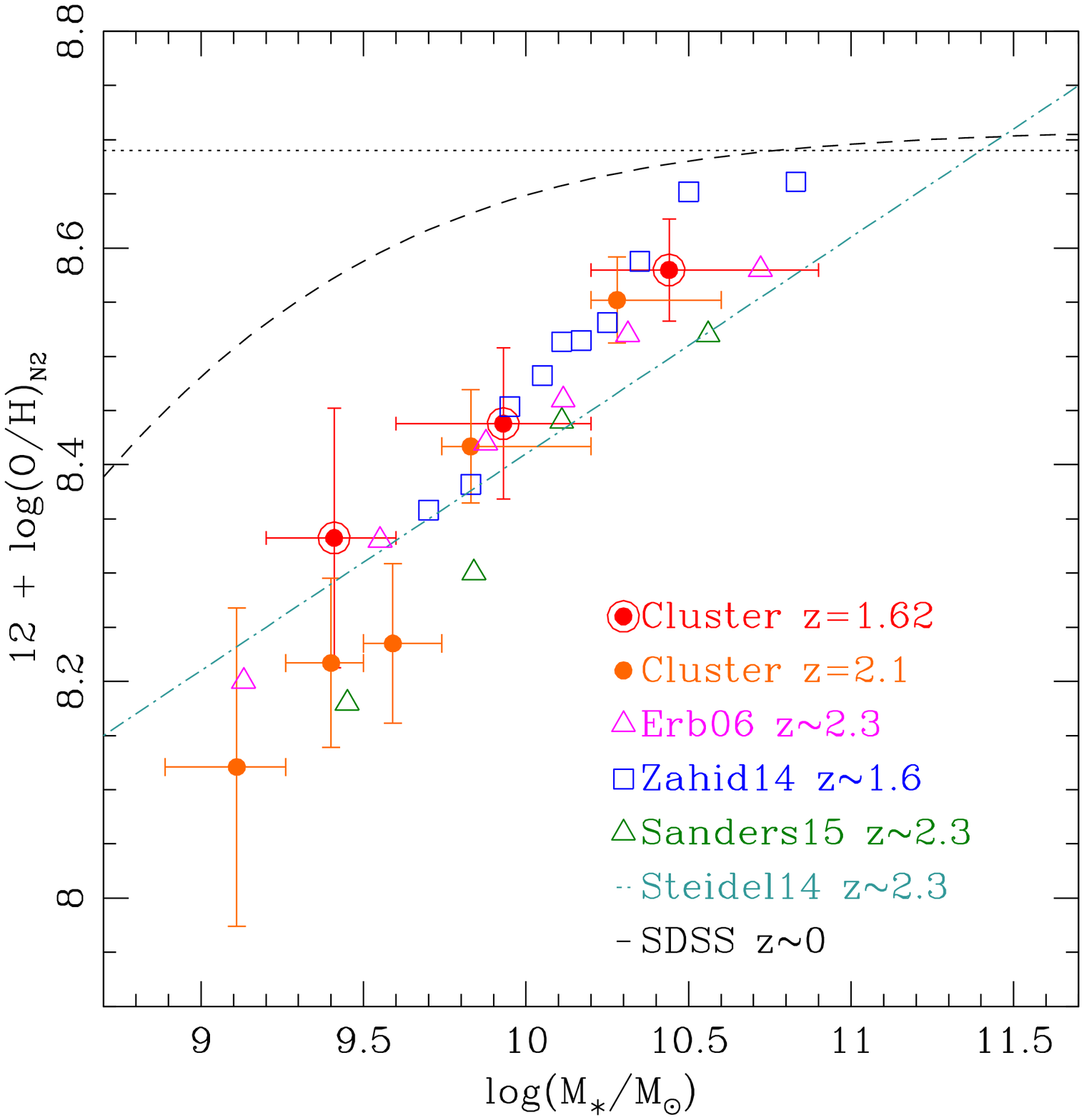}}
\caption{The same as in Fig.~\ref{fig:mzr} but for the stacked spectra
where the cluster galaxies are divided into three mass bins (large red
filled circles; see Fig.~\ref{fig:specstack}).  We show the field data
binned in stellar mass from \citet{zahid:14}, \citet{sanders:15}, and
\citet{erb:06a}.  The \irc\ members are remarkably consistent with the
field MZR relation at $z\sim2$ as well as the MZR measured for the
COSMOS cluster \citep[filled orange circles;][]{kacprzak:15}.
\label{fig:mzr_stack}}
\end{figure*}

The cluster galaxies show increasing metallicity with mass
(Figs.~\ref{fig:mzr} \& \ref{fig:mzr_stack}), a trend that is
consistent with the mass-metallicity relation (MZR) measured for
star-forming galaxies at $z\sim0$ \citep{moustakas:11}.  The cluster
galaxies lie below the local MZR, i.e. the cluster galaxies at
$z\sim1.6$ have lower gas metallicities at a given stellar mass. The
offset in the cluster MZR is similar to that observed for field
star-forming galaxies at $z\sim1.6$ \citep{zahid:14} at $z\sim2.3$
\citep{steidel:14,sanders:15}.



\section{Results}

\subsection{Is \irc\ A Galaxy Cluster?}\label{sec:dispersion}

\irc\ is one of only a handful of galaxy clusters at $z>1.5$ with a
measured cluster velocity dispersion \sigmacl.  The challenge lies in
obtaining medium-resolution near-infrared spectroscopy for a
sufficient number of members to measure accurate redshifts and map the
cluster's kinematic structure (see \ref{sec:kinematics}).  Although
NIR spectroscopy is strongly biased towards emission-line galaxies due
to the faintness of the stellar continuum, this may be advantageous
given the increased star formation at this epoch
\citep[e.g.][]{tran:10,brodwin:13}.  Note that while cluster studies
at $z\sim2$ using HST/WFC3 grism observations can confirm both active
and passive members via, e.g. the D4000 break
\citep{gobat:13,newman:14}, the grisms' redshift accuracy of
$(1+z)\times1000$\kms\ leads to rather uncertain measurements of the
cluster velocity dispersion, e.g. \citet{mei:15}.

Using \Halpha\ to identify 15 emission-line members,
\citet{bayliss:14} measure a very large velocity dispersion of
$1500\pm520$\kms\ in a galaxy cluster at $z=1.48$.  In comparison, the
velocity dispersions of $\sim750-850$~\kms\ measured for two
proto-clusters at $z=2.2$ and $2.5$ \citep{shimakawa:14a} are more in
line with those of lower redshift clusters, e.g. EDisCS
\citep{milvang:08}.  The smallest velocity dispersion observed thus
far at $z\sim2$ is for the COSMOS cluster with $\sigma_{\rm
cl}=552\pm52$~\kms\ \citep{yuan:14}; this cluster also has the most
confirmed members ($>50$) as identified by our ZFIRE survey.  

Relative to these studies, \irc's velocity dispersion of
\sigmacl$=254\pm50$~\kms\ (33 emission-line members;
Fig.~\ref{fig:zhistcl}) is by far the smallest.  This low velocity
dispersion is more in the range of galaxy groups than a galaxy cluster
\citep[e.g.][]{gerke:05,tanaka:13}.  However, this velocity dispersion
is consistent with X-ray results: the cluster velocity dispersion
corresponding to the total cluster mass of
M$_{200}=7.7\pm3.8\times10^{13}$\Msun\ estimated from the weak X-ray
signal detected with XMM-Newton \citep[$2.3\sigma$ level
detection;][]{pierre:12} is $\sigma_{\rm cl, xray}\sim360\pm90$\kms,
$i.e.$ the dynamical velocity dispersion is consistent with the
X-ray derived estimate at $\sim1.5\sigma$.  These values also are
consistent with predictions from numerical simulations, $e.g.$
\citet{evrard:08} show that a halo with X-ray mass of
M$_{200}\sim7.7\times10^{13}$\Msun\ has an estimated velocity
dispersion of 400\kms.  Note that the cluster members are spatially
concentrated with 20 of the 33 members within \Rproj$=1$~Mpc of the
BCG (Fig.~\ref{fig:zhistcl}, right).

Using the Millenium simulation with a semi-analytic galaxy formation
model, \citet{shattow:13} select 50 central galaxies in halos with
masses greater than $7\times10^{13}$\Msun\ and track them from $z=6$
to $z=0$; note that this halo mass is essentially \irc's dynamical
mass as measured by both kinematics and X-ray.  The study finds that
by $z=0$, most of the halos have masses $>2\times10^{14}$\Msun,
i.e. they are likely to be clusters if not the most massive ones
\citep[see also][]{chiang:13}.  Thus \irc\ will evolve into a galaxy
cluster that can be as massive as the Virgo cluster
\citep{schindler:99} and possibly as massive as $10^{15}$\Msun.

\subsection{Nebular vs. Stellar Attenuation}

\EBVgas\ and \EBVsed\ can be used to determine dust properties, $i.e.$
\EBVgas\ probes sightlines to regions of ionized gas that are heated
by short-lived OB stars while \EBVsed\ is a measure of the (UV) dust
attenuation of the stars and integrated over longer timescales ($>100$
Myr).  For local starbursts, \EBVsed\ and \EBVgas\ are correlated such
that the color excess of the stellar continuum is about half that of
the ionized gas \citep[$f=0.44$;][]{calzetti:00}, i.e. the attenuation
as measured by the nebular lines is larger.  It is unclear if this
relation holds at $z>1$ with studies measuring ratios varying from
$\sim0.26-1$
\citep{erb:06b,yoshikawa:10,kashino:13,zahid:14,price:14,reddy:15}.

We are able to directly measure the color excess \EBVgas\ for 9
cluster galaxies where \Halpha/\Hbeta$>2.86$ and
\Hbeta$\geq0.5$\ergu\ is detected at $\geq3\sigma$
(Fig.~\ref{fig:ebv}, right).  Both \EBVgas\ and \EBVsed\ span ranges
($0-0.8$) that are comparable to values measured at $z\sim1-2$ in both
field and cluster studies \citep{zeimann:13,price:14}.  The ratio of
nebular to stellar attenuation is also consistent with the measured
values of $\sim0.44$ \citep{calzetti:00,yoshikawa:10}, albeit with
considerable scatter.

There is no clear trend between the nebular attenuation \Agas\ and
stellar mass (Fig.~\ref{fig:Av_smass}), nor is there any indication
that nebular attenuation depends on environment as suggested by past
studies \citep{koyama:13b,zeimann:13}.  In contrast, the stellar
attenuation \ACalsed\ (starburst, \Rv=4.05) does increase with stellar
mass (Fig.~\ref{fig:Av_smass}) but with a slope that is shallower than
observed in the field at $z\sim1.5$ \citep{price:14}.  Our results
suggest that the attenuation measured by the integrated stellar light
may be less affected by, e.g. patchy dust distribution; however, our
sample size is limited.  We refer the reader to
\citet{reddy:15} for a more extensive study of dust based on a
significantly larger sample of (field) galaxies at $z\sim2$.

\subsection{Star Formation in the Cluster Core}

With our sensitive spectroscopic survey of \irc, we measure
extinction-corrected \Halphased\ SFRs for 26 individual members with
$\sim2-50$\Msunyr; AGN have been removed.  Note that an advantage of
spectroscopy over using narrow-band imaging to derive
\Halphased\ SFRs is that we can remove AGN using \NII/\Halpha\ as
well as those detected using X-ray.  With the 26 cluster galaxies, the
lower limit on \irc's integrated \Halphased\ star formation rate is
$\sim325$\Msunyr\ (Table~\ref{tab:SFR}).  The higher integrated
\Halphased\ SFR in the cluster core (Fig.~\ref{fig:ISFR}) supports our
earlier result based on \mipsmu\ observations of \irc\
\citep{tran:10}.

A handful of studies have now also obtained \Halpha\ spectroscopy of
cluster galaxies at $z\sim1.5$ using ground-based telescopes and the
WFC3 grism in \hubble.  Line sensitivities vary, but the consensus is
that most galaxy clusters in this redshift range have integrated SFRs
within \Rproj$<0.5$~Mpc of up to $\sim200$\Msunyr\ and SFRs of
individual members can be as high as $100$\Msunyr\
\citep{hayashi:11,zeimann:13}.  With an \Halphased\ SFR of
$\sim60$\Msunyr\ integrated within \Rproj$<0.5$~Mpc
(Table~\ref{tab:SFR}), \irc\ is consistent with these measurements.


Our analysis uses the integrated \Halphased\ SFRs within different
\Rproj\ (Fig.~\ref{fig:ISFR}; Table~\ref{tab:SFR}) to enable direct
comparison to the literature.  Note that the integrated
\Halphased\ SFR within \Rproj\ is different from the {\it fraction} of
star-forming members.  Because the overall galaxy density is higher in
clusters, it is possible to have both a large integrated SFR and an
increase in quiescent galaxies relative to the field,
e.g. \citet{quadri:12} and \citet{santos:14}.  For example, the
\Halphased-SFR per square Mpc is highest in the core
(\Rproj$<0.25$~Mpc), but the average SFR per galaxy is only about half
that of galaxies at \Rproj$<2$~Mpc (Fig.~\ref{fig:ISFR}).

Our results confirm that the integrated \Halphased\ SFRs in the
cores of galaxy clusters at $z\gtrsim1.5$ are significantly higher
than in massive galaxy clusters at $z<1.4$
\citep{kodama:04,finn:05,bauer:11}.  The astute reader will notice
that our \Halphased\ SFRs are lower than the \mipsmu\ values published
in \citet{tran:10}.  For the six \irc\ members with both \Halphased\
and \mipsmu\ SFRs, the average ratio is $\sim2.9\pm1.1$, $i.e.$ the
\mipsmu\ SFRs are $\sim3$ times higher.  However, it is likely that
the \mipsmu\ SFRs are over-estimated: \citet{papovich:07} showed that
for galaxies at $z\sim2$, SFRs based on \citet{chary:01} templates
fitted only to \mipsmu\ are $\sim2-5$ higher compared to SFRs
determined with \mipsmu, $70\mu$m, and $160\mu$m.

We note that if we use the \mipsmu\ measurements for the six
\Halpha-detected members, the the number of cluster galaxies with
\Halpha\ SFRs$>30$\Msunyr\ increases from two to six.  Because we are
also constrained by the spatial sampling of the NIR spectroscopy, we
consider the \Halphased\ SFRs in Table~\ref{tab:SFR} to be lower
limits.  However, this result is tentative and a larger cluster sample
at $z\sim2$ is needed to better quantify the relation between \Halpha\
and \mipsmu, especially for SFR$>20$\Msunyr.


\subsection{Star Formation Rate versus Stellar Mass}\label{sec:sfr_smass}

The 26 \Halpha-detected cluster galaxies have stellar masses of
$\sim10^9-10^{11}$\Msun\ and show a trend of increasing SFR with
stellar mass that is similar to the field (Fig.~\ref{fig:sfr_smass},
left), i.e. a ``main sequence'' of star-forming members
\citep{daddi:07,koyama:13b}.  Most of the lower-mass cluster galaxies
(\Mstar$<10^{10}$\Msun) will more than double their stellar masses
given the elapsed cosmic time since $z=1.62$ if they maintain
their current Specific SFRs (SSFR; see also Table~\ref{tab:SFR}).
Our results are consistent with \citet{zeimann:13} who use WFC3 grism
measurements of \Halpha\ to show that cluster galaxies at $1<z<1.5$
follow a SFR-\Mstar\ relation.


To determine if the SFR-\Mstar\ relation depends on environment, we
use the Four-Star Galaxy Evolution (ZFOURGE) survey \citep[Straatman
et al., submitted]{tomczak:14} to select star-forming field galaxies
at $1.5<z<2.0$ with SFRs measured by combining UV and mid-IR fluxes
Tomczak et al., submitted).  Given our
\Halphased\ detection limit, we apply a SFR detection threshold of
2\Msunyr\ to both the \irc\ and ZFOURGE samples.  We compare
\Halphased\ SFRs and the specific star formation rate (SSFR defined to
be SFR/\Mstar) to stellar mass in Fig.~\ref{fig:sfr_smass}.  We
also compare to recent results from MOSDEF based on \Halpha\
measurements of field galaxies at $z\sim2.3$ \citep{sanders:15}. 

The \irc\ members have systematically lower SFRs compared to the
field.  However, the difference can be due to using two different
methods of measuring SFRs \citep[\Halphased\
vs. UV+MIR;][]{kennicutt:12}.  Instead, we consider the slope to test
for any environmental difference.  By fitting a least-squares to \irc\
and bootstrapping the sample ($10^4$ realizations), we measure the
SSFR-\Mstar\ slope to be $-0.51\pm0.30$.  For the ZFOURGE field
galaxies, we measure a shallower slope of $-0.23\pm0.07$.  The
difference in the SSFR-\Mstar\ slope between
\irc\ and ZFOURGE suggests that environment plays a role, but our
results are not statistically significant.

An alternate comparison is to \citet{sanders:15} who measure
\Halpha\ SFRs for the field at $z\sim2.3$ (Fig.~\ref{fig:sfr_smass}.
The \irc\ members are clearly offset to lower SFRs compared to the
\Halpha-detected field galaxies.  {One possible explanation for
the lower SFRs in \irc\ is the elapsed cosmic time between $z\sim2.3$
and $z\sim1.6$ ($\Delta t\sim1.1$~Gyr).  However, the offset in the
SFR-\Mstar\ relation for MOSDEF field galaxies at $z\sim1.6$ at
$z\sim2.3$ \citep{kriek:15} is not as significant as observed in \irc.
We conclude that our observations indicate a dependence on
environment, i.e. lower SFRs in the cluster especially for the higher
mass members.

\subsection{Gas Phase Metallicity-Mass Relation (MZR)}

There is considerable and confusing disagreement over the observed
relation between gas phase metallicity and stellar mass (MZR) at
$z\sim2$.  While several studies of field and protocluster galaxies
find that metallicities increase with stellar mass
\citep{zahid:14,steidel:14,shimakawa:15}, others do not
\citep{stott:13,kulas:13}.  Using the N2 method calibrated by
\citet{pettini:04}, we measure the oxygen abundance in cluster
galaxies and compare to their stellar masses (Figs.~\ref{fig:mzr}
\& \ref{fig:mzr_stack}).  We find that the \irc\ members are
offset to lower metallicities relative to star-forming galaxies in the
local universe \citep{tremonti:04}.  

The cluster MZR is remarkably similar to that of several field surveys
at $z\sim2$ \citep{erb:06a,steidel:14,sanders:15,zahid:14}; this is
particularly evident when comparing the stacked spectra
(Fig.~\ref{fig:mzr_stack}).  To quantify the trend of increasing
metallicity with stellar mass, we fit a linear least-squares to the 20
members with measured \NII/\Halpha\ (we exclude the five members with
only upper limits on \NII):

\begin{equation}
12 + \log {\rm (O/H)}_{\rm N2} = 8.49 + 0.13 [\log({\rm
M}_{\star}/{\rm M}_{\odot}) - 10.0 ]
\end{equation}

where $\sigma_{\rm RMS}=0.10$.  The MZR in \irc\ is comparable to the
relation measured by \citet{steidel:14} for field galaxies at
$z\sim2.3$ where the slope and normalization are $(0.2,8.41)$.  The
consistency holds across the full range in mass
[$9\lesssim$\logMstarMsun$\lesssim11$].  For these 20 members, the
median difference between measured \NII\ and that predicted using the
fit from \citet{zahid:14} for field galaxies at $z\sim1.6$ is only
$0.04$.

Whether there is an environmental imprint on the MZR at $z\sim2$ is
yet to be clearly established.  \citet{ellison:09} detect a small
offset of $\sim0.04$ dex to higher metallicities in overdense
environments at $z\sim0$ \citep[see also][]{cooper:08b,peng:14}, and
recent studies at $z\sim2$ claim to find a measurable offset between
their cluster and field samples
\citep{valentino:14,shimakawa:15}. However, we find no evidence of an 
environmental dependence for the MZR when comparing \irc\ to the field
at $z\sim1.6$.

What is striking is how the cluster MZR does not evolve at $1.6<z<2.1$
(over $\sim1$ Gyr): \irc's MZR is virtually identical to the COSMOS
cluster at $z=2.1$
\citep[Fig.~\ref{fig:mzr_stack};][]{kacprzak:15}.  The similarity of
the cluster MZRs matches the essentially constant gas metallicities of
lensed field galaxies at $1.5<z<2.5$ \citep{yuan:13}.  Our results are
also in line with the slow evolution of the MZR predicted by
simulations \citep{dave:11b}, but we note that these same simulations
also predict an environmental imprint, e.g. higher metallicities at a
given stellar mass, that we do not observe in either galaxy cluster.

Our results show that the cluster MZR and its evolution is
indistinguishable from that of the field at $1.6<z<2.3$ which is
consistent with predictions from recent simulations by
\citet{taylor:15}.  We conclude that the somewhat contradictory
results observed thus far for the MZR at $z\sim2$ are due to a
combination of different line flux sensitivities and sample sizes of
the various studies.  By expanding our sensitive survey of cluster
galaxies at $z\sim2$ and comparing to results from MOSDEF
\citep{kriek:15}, we hope to resolve these disagreements.

\subsection{Star Formation Rates and the MZR}

Studies at low redshift find that at a given stellar mass, the gas
metallicity is lower for galaxies with higher star formation rates
\citep{mannucci:10,andrews:13}.  The Fundamental Mass-Metallicity
relation (FMR) quantifies this dependence of the MZR on the SFR and
can be explained by a SFR dependence on the ratio of self-enrichment
versus inflow of pristine gas. Whether the FMR holds at $z\sim2$ is
unclear with recent studies finding no evidence of an MZR dependence
on SFR \citep{wuyts:14,sanders:15}.  Evidence for an environmental
dependence for the FMR at $z\sim2.3$ is also tentative at best
\citep{shimakawa:15}.

To test for a dependence on SFR, we divide the \irc\ members into
approximately equal sized low and high SFR bins
(SFR[\Halphased$]=10$\Msunyr; Fig.~\ref{fig:mzr}); we exclude the five
members that have only upper limits on their gas metallicities because
their true values could be considerably lower.  At high stellar masses
[\logMstarMsun$>10$], the high SF members dominate with a metallicity
range of $\sim0.2$~dex. At low stellar masses, the low SF members
dominate with a larger metallicity range of $\sim0.4$~dex.

We fit a least-squares to the 11 high SFR members and the 9 low SFR
members with measured \NII\ and measure slopes with errors of
$0.05\pm0.08$ and $0.10\pm0.11$ respectively.  The MZR for both low
and high SFR bins are consistent (within $1\sigma$) with that of the
cluster sample as a whole as well as the MZR measured by
\citet{steidel:14}.  We conclude that the \irc\ members show no
evidence of an MZR dependence on the SFR, but we acknowledge that
our cluster sample is limited.


\section{Evidence of Environmental Dependence?}

With our NIR spectroscopy, we measure \Halpha\ and \NII\ fluxes that,
combined with our SED fits from FAST, provides extinction-corrected
\Halpha\ star formation rates, gas phase metallicities from
\NII/\Halpha, and stellar masses.  The primary evidence we find for
any environmental dependence is in the cluster star formation rates.
\irc's integrated \Halphased\ SFR per unit area is about a factor of
three higher at \Rproj$<0.25$~Mpc compared to \Rproj$<1$~Mpc (from 144
to 26\Msunyr~Mpc$^{-2}$; Table~\ref{tab:SFR}).  This is driven by the
higher number density of star-forming galaxies in the cluster core
relative to studies of cluster cores at lower redshifts
\citep{tran:07,mei:09}.  However, the average \Halphased\ SFR per
galaxy is actually lower in the core compared to the values at
\Rproj$=0.5$ and 1 Mpc \citep[Table~\ref{tab:SFR}; see
also][]{brodwin:13}. 

At a given stellar mass, the \irc\ members have \Halphased\ SFRs that
are lower compared to the field at $z\sim2.3$
(Fig.~\ref{fig:sfr_smass}).  The slope of the SFR-\Mstar\ relation
also seems to depend on environment such that the massive cluster
galaxies have lower SFRs compared to their field counterparts.  The
quenching of SF in the massive cluster galaxies at $z\sim1.6$ means
these members will passively age to match their cluster counterparts
at $z<1$ \citep{mei:09}.  The continued stellar growth for lower mass
cluster galaxies is also consistent with the more extended star
formation histories measured in cluster galaxies at $z<1$
\citep{tran:05a,tran:07}.

Like field galaxies at $z\sim1.5$, the cluster galaxies show an
increasing amount of attenuation with stellar mass, $i.e.$ more
massive galaxies are dustier.  This trends hold primarily for the
stellar attenuation (\ACalsed) as measured by the SED fitting to the
multi-wavelength photometry (Fig.~\ref{fig:Av_smass}).  However, the
large scatter in both in \irc\ and field studies
\citep{yoshikawa:10,price:14} is sobering, particular when comparing
the nebular attenuation \Agas\ to \Mstar.  It is very likely that the
empirical relations used to determine attenuation and thus obtain
extinction-corrected SFRs at $z\sim2$ have uncomfortably large
uncertainties and depend on SFRs \citep[e.g.][]{reddy:15}.  A better
characterization of dust attenuation laws for difference spectral
types at $z>1$ using essentially low resolution spectroscopy
\citep{kriek:13} holds promise in addressing this issue.

The gas phase metallicity versus \Mstar\ (MZR) for \irc\ members is
nearly identical to the field at $z\sim1.6$ (Figs.~\ref{fig:mzr} \&
\ref{fig:mzr_stack}).  The cluster MZR does not depend on SFR which is
also consistent with field results
\citep{wuyts:14,steidel:14,sanders:15}.  Comparing to our results for
the COSMOS cluster at $z=2.1$ \citep{kacprzak:15} shows no evolution
in the MZR in the $\sim1$~Gyr.  Our ZFIRE cluster results contradict
recent claims at $z\sim2$ of a measurable difference in the MZR with
environment \citep{kulas:13,shimakawa:15}.  We note that the
environmental imprint on the MZR is small even using SDSS at $z\sim0$
\citep{cooper:08b,ellison:09,peng:14}, thus if there is an
environmental dependence for the MZR at $z\sim2$, a considerably
larger sample of cluster galaxies is needed to detect it.  

Alternatively, calibration issues with using \NII/\Halpha\ to measure
gas phase metallicities at $z\sim2$ may mask any environmental imprint
on the MZR.  {For example, \citet{kewley:08} showed that the
choice of line ratios used to measure the gas-phase
metallicity has a significant effect on the shape of the MZR.  By
obtaining the full suite of emission lines for the BPT diagram
\citep{baldwin:81} to separate star-forming galaxies from those
hosting AGN, we can compare to metallicities using alternative line
ratios such as \Hbeta/\OIII\ \citep[e.g.][]{shapley:15}.  We can also
characterize the ionization conditions.  If there is a relationship
between SSFR and ionization parameter, this would have interesting
implications for the properties of star clusters at high redshift,
e.g. the spatial distribution and masses of stars embedded within them
\citep{kewley:15}.  By expanding our survey of cluster galaxies at
$z\sim2$, we will explore how gas metallicities and ionization
parameters are linked to star formation, galaxy (stellar) mass, and
environment.

\section{Conclusions}

We survey the galaxy cluster \irc\ using the multi-object
spectrographs LRIS (optical) and MOSFIRE (near-IR) on Keck I.  We
obtain 130 unique redshifts based mostly on emission lines, e.g. \OII\
and \Halpha, and identify 45 galaxies with $1.58<$\zspec$<1.8$
(Figs.~\ref{fig:zhist} \& \ref{fig:zhistcl}).  The
\irc\ members define a narrow redshift distribution with 33 galaxies
at $1.6118<$\zspec$<1.6348$.  The corresponding cluster redshift and
velocity dispersion are $z_{\rm cl}=1.6233\pm0.0003$ and $\sigma_{\rm
cl}=254\pm50$\kms.  

While \irc's velocity dispersion is the lowest measured thus far for a
cluster at $z\sim2$, it is consistent with the $\sigma$ inferred from
the cluster mass estimate of M$_{200}=7.7\pm3.8\times10^{13}$\Msun\
from the weak X-ray detection \citep{pierre:12}.  Simulations also
show that most systems with this velocity dispersion and redshift will
evolve into a cluster by $z\sim0$.  \irc\ is spatially concentrated
with 20 spectroscopically confirmed members within \Rproj$<1$~Mpc of
the BCG.  The cluster's high spatial concentration and low velocity
dispersion are conducive to galaxy-galaxy merging
\citep{rudnick:12,lotz:13}.

With MOSFIRE, we reach near-IR line flux sensitivities of 0.3\ergu\
that, combined with our SED fits to multi-wavelength photometry,
provide extinction-corrected \Halpha\ star formation rates, gas phase
metallicities using \NII/\Halpha, and stellar masses.  Adopting a
\Halphased\ SFR limit of 2\Msunyr\ at \zcl\ and removing AGN, we
confirm our earlier result based on \mipsmu\ observations of elevated
star formation in \irc's core \citep{tran:10}: the integrated SFR per
unit area at \Rproj$<0.25$~Mpc is about a factor of three larger than
at \Rproj$<1$~Mpc (Fig~\ref{fig:ISFR}; Table~\ref{tab:SFR}).  We
consider the \Halphased\ SFRs to be lower limits due to sparse
sampling and because for the handful of objects where we also can
measure the Balmer decrement and/or \mipsmu\ flux, the individual SFRs
increase by $\sim2-3$.

In our analysis, the strongest evidence for any environmental
dependence at $z\sim1.6$ is in the cluster star formation rates.  The
high \Halphased\ SFR in the core (\Rproj$<0.25$~Mpc;
Fig.~\ref{fig:ISFR}) is driven by the concentration of star-forming
members while the average SFR per galaxy in the core is actually half
that of galaxies at \Rproj$\sim1$~Mpc (Table~\ref{tab:SFR}).  The
massive cluster galaxies also tend to have lower SFRs compared to
their field counterparts (Fig.~\ref{fig:sfr_smass}).  This
mass-dependent quenching of SF is needed for these cluster galaxies to
match the stellar ages of their cluster counterparts at $z<1$
\citep[$e.g.$][]{tran:07,mei:09}.

Like the field, the \irc\ members show a trend of increasing
attenuation with stellar mass as measured by the SED fits (\ACalsed),
$i.e.$ more massive galaxies are dustier (Fig.~\ref{fig:Av_smass}).
However, this is not the case when using the nebular attenuation as
measured by the Balmer decrement (\Agas).  The significant scatter in
the nebular attenuation suggests larger problems with correctly
measuring dust attenuation at $z\sim2$ for both cluster and field
galaxies \cite[see also][]{reddy:15}.

The gas phase metallicity versus stellar mass (MZR) for the cluster
galaxies is virtually identical to the field at $z\sim1.6$, and both
are offset to lower metallicities compared to galaxies at $z\sim0$
(Figs.~\ref{fig:mzr} \& \ref{fig:mzr_stack}).  Comparing to the fit
determined by \citet{zahid:14} for field galaxies at $z\sim1.6$, the
median difference between measured and predicted \NII\ is only $0.04$.
Fitting a least squares to the \irc\ members confirms that their MZR
is the same as for the field as measured by, $e.g.$ \citet{steidel:14}
at $z\sim2.3$.  The cluster MZR does not depend on star formation rate
which is also consistent with recent results on field galaxies
\citep{wuyts:14,sanders:15}.

Particularly striking is that \irc\ and the COSMOS cluster at $z=2.1$
\citep{kacprzak:15} have the same MZR even though the elapsed time is
$\sim1$~Gyr.  Both clusters' MZRs are also well-matched to the field
as measured by several surveys at $z\sim2$.  We do not see any
evidence of an environmental imprint on the MZR at $z\sim2$ as
indicated by previous studies \citep{kulas:13, shimakawa:15}.  We
attribute this difference to our combination of a larger cluster
sample with a more sensitive line flux limit.

In summary, environmental effects are likely to be subtle at $z\sim2$
and require a larger survey of cluster galaxies to fully characterize.
With ZFIRE, we are building such a cluster sample and also expanding
to include additional line diagnostics needed to test, $e.g.$ the
reliability of \NII/\Halpha\ for measuring gas phase metallicity.
ZFIRE complements ongoing NIR spectroscopic surveys of field galaxies
at $z\sim2$, $e.g.$ KBSS \citep{steidel:14}, MOSDEF
\citep{kriek:15}, and KMOS$^{\rm 3D}$ \citep{wisnioski:15}, and will
enable us to test for evolution in spatially resolved quantities as a
function of environment.

\acknowledgements

We are grateful to the MOSFIRE team and in particular N. Konidaris for
his help throughout the data reduction process.  The Keck support
staff were equally essential in gathering and processing the MOSFIRE
observations with special thanks to M. Kassis, J. Lyke, and G. Wirth.
K. Tran thanks M. Kriek, A. Shapley, S. Price, and D. Calzetti for
helpful discussions, and R. Sanders for use of their data.  This work
was supported by a NASA Keck PI Data Award administered by the NASA
Exoplanet Science Institute. Data presented herein were obtained at
the W. M. Keck Observatory from telescope time allocated to NASA
through the agency's scientific partnership with the California
Institute of Technology and the University of California. The
Observatory was made possible by the generous financial support of the
W. M. Keck Foundation.  This material is based upon work
supported by HST GO program \#12896 which was provided by NASA through
a grant from the Space Telescope Science Institute, which is operated
by the Association of Universities for Research in Astronomy, Inc.,
under NASA contract NAS 5-26555.  This material is also based upon
work supported by the National Science Foundation under Grant
\#1410728.  GGK acknowledges the support of the Australian Research
Council through the award of a Future Fellowship (FT140100933).  The
authors wish to recognize and acknowledge the very significant
cultural role and reverence that the summit of Mauna Kea has always
had within the indigenous Hawaiian community. We are most fortunate to
have the opportunity to conduct observations from this mountain.

\bibliographystyle{/Users/vy/Work/aastex/apj}
\bibliography{/Users/vy/Work/aastex/tran}


\end{document}